\documentstyle[12pt,epsfig]{article}
\def\be{\begin{equation}}
\def\ee{\end{equation}}
\def\bc{\begin{center}}
\def\ec{\end{center}}
\def\bea{\begin{eqnarray}}
\def\eea{\end{eqnarray}}
\def\dd{\displaystyle}

\def\ov{\overline}
\def\gev{{\rm \; GeV}}

\def\str{{\tilde{t}_R}}
\def\stl{{\tilde{t}_L}}

\def\mst{{m_{\tilde{t}}}}
\def\mch{{m_{\tilde{\chi}}}}
\def\simlt{\stackrel{<}{{}_\sim}}
\def\simgt{\stackrel{>}{{}_\sim}}
\def\tb{{\tan \beta}}

\catcode`@=11
\def\marginnote#1{}
\newcount\hour
\newcount\minute
\newtoks\amorpm
\hour=\time\divide\hour by60
\minute=\time{\multiply\hour by60 \global\advance\minute by-\hour}
\edef\standardtime{{\ifnum\hour<12 \global\amorpm={am}%
        \else\global\amorpm={pm}\advance\hour by-12 \fi
        \ifnum\hour=0 \hour=12 \fi
        \number\hour:\ifnum\minute<10 0\fi\number\minute\the\amorpm}}
\edef\militarytime{\number\hour:\ifnum\minute<10 0\fi\number\minute}
\def\draftlabel#1{{\@bsphack\if@filesw {\let\thepage\relax
   \xdef\@gtempa{\write\@auxout{\string
      \newlabel{#1}{{\@currentlabel}{\thepage}}}}}\@gtempa
   \if@nobreak \ifvmode\nobreak\fi\fi\fi\@esphack}
        \gdef\@eqnlabel{#1}}
\def\@eqnlabel{}
\def\@vacuum{}
\def\draftmarginnote#1{\marginpar{\raggedright\scriptsize\tt#1}}
\def\draft{\oddsidemargin 0.0truein
        \def\@oddfoot{\sl preliminary draft \hfil
        \rm\thepage\hfil\sl\today\quad\militarytime}
        \let\@evenfoot\@oddfoot \overfullrule 3pt
        \let\label=\draftlabel
        \let\marginnote=\draftmarginnote
   \def\@eqnnum{(\theequation)\rlap{\kern\marginparsep\tt\@eqnlabel}%
\global\let\@eqnlabel\@vacuum}  }
\catcode`@=12
\begin{document}
\begin{titlepage}
\vspace*{-1cm}
\phantom{bla}
\hfill{CERN-TH/95-340}
\\
\phantom{bla}
\hfill{DFPD~95/TH/66}
\\
\phantom{bla}
\hfill{hep-ph/9601293}
\vskip 1.5cm
\begin{center}
{\Large\bf Phenomenological implications
\\
\vskip .2cm
of light stop and higgsinos}
\footnote{Work supported in part by the European Union under
contract No.~CHRX-CT92-0004.}
\end{center}
\vskip 1.0cm
\begin{center}
{\large Andrea Brignole}
\\
\vskip .1cm
Theory Division, CERN, CH-1211 Geneva 23, Switzerland
\\
\vskip .2cm
{\large Ferruccio Feruglio}
\\
\vskip .1cm
Dipartimento di Fisica, Universit\`a di Padova, I-35131 Padua, Italy
\\
\vskip .2cm
and
\\
\vskip .2cm
{\large Fabio
Zwirner}\footnote{On leave from INFN, Sezione di Padova, Padua,
Italy.}
\\
\vskip .1cm
Theory Division, CERN, CH-1211 Geneva 23, Switzerland
\end{center}
\vskip 0.5cm
\begin{abstract}
\noindent
We examine the phenomenological implications of light
$\str$ and higgsinos in the Minimal Supersymmetric
Standard Model, assuming $\tan^2 \beta < m_t / m_b$
and heavy $\stl$ and gauginos. In this simplified
setting, we study the contributions to $\Delta m_{B_d}$,
$\epsilon_K$, $BR(b \rightarrow s \gamma)$, $R_b \equiv
\Gamma (Z \rightarrow b \overline{b}) / \Gamma ( Z
\rightarrow {\rm hadrons})$, $BR(t \to b W)$, and their
interplay.
\end{abstract}
\vfill{
CERN-TH/95-340
\newline
\noindent
December 1995}
\end{titlepage}
\setcounter{footnote}{0}
\vskip2truecm
\vspace{1cm}
{\bf 1.}
If low-energy supersymmetry (for a review and references,
see e.g. \cite{sergio}) plays a role in the resolution of the
naturalness problem of the Standard Model (SM), then the Minimal
Supersymmetric Standard Model (MSSM) is the most plausible effective
theory at the electroweak scale, and we should be close to the
discovery of Higgs bosons and supersymmetric particles. The scalar
partners of the top quark (two complex spin-0 fields, one for each
chirality of the corresponding quark) and the fermionic partners of
the gauge and Higgs bosons (two charged Dirac particles, or
charginos, and four neutral Majorana particles, or neutralinos)
are among the most likely candidates for an early discovery.
A particularly attractive possibility is the existence of light
$\str$ and higgsinos, within the discovery reach of LEP2, as
suggested by some MSSM fits to precision electroweak data
(see \cite{fits} for the different points of view). The aim
of the present paper is to elucidate the phenomenological
implications of such a possibility, developing some of the
observations already present in \cite{fits} in a more systematic
and transparent way.

In the rest of this section, we present simplified expressions
for the light sparticle masses in the limit of interest.
In section~2, we introduce simplified MSSM analytical formulae
for a number of physical observables, such as $\Delta m_{B_d}$,
$\epsilon_K$, $BR ( b \rightarrow s \, \gamma)$, $R_b \equiv
\Gamma (Z \rightarrow b \overline{b}) / \Gamma ( Z \rightarrow
{\rm hadrons})$, $BR(t \to b W)$. For each class of processes, we
discuss how the existing experimental data constrain the MSSM
parameter space. In section~3, we present our conclusions.

Assuming that the squark mass matrices can be diagonalized
(in generation space) simultaneously with those of the
corresponding quarks, the spectrum of the stop sector
is described, in the usual MSSM notation and in the
$(\tilde{t}_L,\tilde{t}_R)$ basis, by the $2 \times 2$ matrix
\be
\begin{array}{c}
{\cal M}^2_{\tilde{t}}
 =
\left(
\begin{array}{cc}
m_{LL}^2 & m_{LR}^2
\\
m_{LR}^2 & m_{RR}^2
\end{array}
\right)
\\ \\
=
\left(
\begin{array}{cc}
m_{Q_3}^2 + m_t^2 + {1 \over 6} (4 m_W^2 - m_Z^2) \cos 2 \beta
& m_t ( A_t - \mu \cot \beta )
\\
m_t ( A_t - \mu \cot \beta ) &
m_{U_3}^2 + m_t^2 + {2 \over 3} (- m_W^2 + m_Z^2) \cos 2 \beta
\end{array}
\right)
\, .
\end{array}
\ee
In the limit $m_{RR}^2 , m^2_{LR} \ll m_{LL}^2$, the lightest stop
eigenstate is predominantly a $\str$, $\tilde{t}_2 = - \sin
\theta_t \tilde{t}_L + \cos \theta_t \tilde{t}_R$, with $\theta_t
\simeq m_{LR}^2/m_{LL}^2 \ll 1$, and the corresponding mass
eigenvalue is given by $m_{\tilde{t}_2}^2 \simeq m_{RR}^2 -
(m_{LR}^2)^2 / m_{LL}^2$. The above situation could arise for
example when $\tb < m_t/m_b$, or $h_t \gg h_b$, since in that
case the structure of the renormalization group equations favours
$m_{U_3} < m_{Q_3} < m_{\tilde{q}}$, where $m_{\tilde{q}}$ is some
average squark mass. It is also known \cite{fits} that it is
easier to reconcile a light $\str$  than a light $\stl$ with the
stringent limits on the effective $\rho$ parameter coming from
the electroweak precision data .

\begin{figure}[h]
\vspace{-0.1cm}
\centerline{
\epsfig{figure=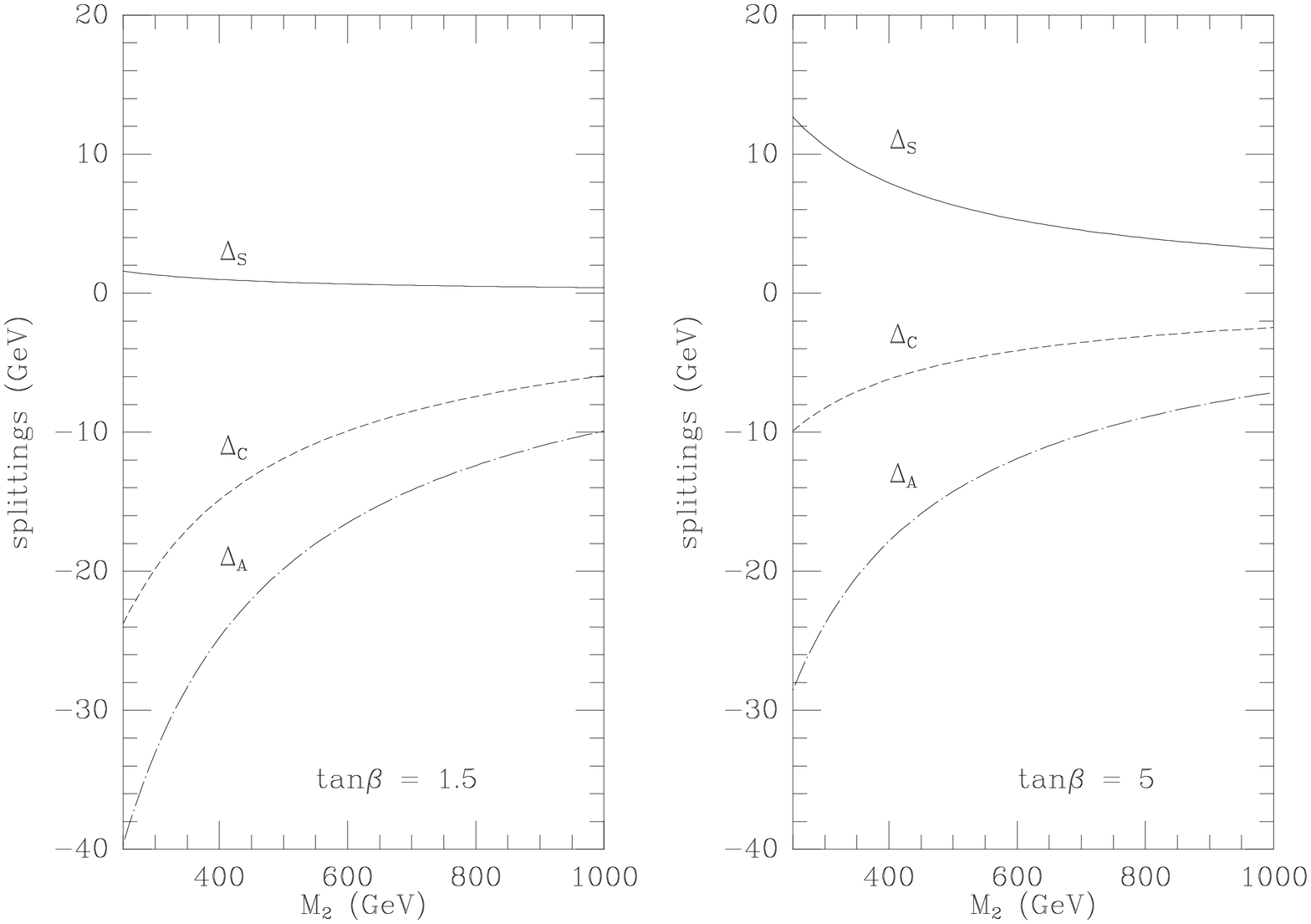,height=9cm,angle=-90}
}
\vspace{-0.3cm}
\caption{\it The mass splittings $\Delta_S$ (solid lines),
$\Delta_A$ (dash-dotted lines) and $\Delta_C$ (dashed lines), as
functions of $M_2$, for the representative values $\tb=1.5,5$.}
\label{sample}
\end{figure}

The mass matrices in the chargino and neutralino sector read
\be
{\cal M}_C =
\left(
\begin{array}{cc}
M_2 & \sqrt{2} m_W \sin \beta \\
\sqrt{2} m_W \cos \beta & \mu
\end{array}
\right) \, ,
\ee
and
\begin{equation}
{\cal M}_N =
\left(
\begin{array}{cccc}
M_1 & 0 & - m_Z c_{\beta} s_W & m_Z s_{\beta} s_W \\
0 & M_2 &   m_Z c_{\beta} c_W & - m_Z s_{\beta} c_W \\
- m_Z c_{\beta} s_W & m_Z c_{\beta} c_W & 0 & - \mu  \\
 m_Z s_{\beta} s_W & - m_Z s_{\beta} c_W & - \mu & 0  \\
\end{array}
\right) \, ,
\end{equation}
where ${\cal M}_N$ has been written in the $(-i \tilde{B},
-i \tilde{W}_3, \tilde{H}_1^0, \tilde{H}_2^0)$ basis, and
$s_{\beta} \equiv \sin \beta$, $c_{\beta} \equiv \cos \beta$,
$s_W \equiv \sin \theta_W$, $c_W \equiv \cos \theta_W$. An
approximate Peccei-Quinn symmetry, recovered in the limit
$\mu \rightarrow 0$, may originate the hierarchy $\mu \ll
M_1, M_2$, which leads to one charged and two neutral
higgsinos much lighter than the other mass eigenstates.
In first approximation, we find the following three
degenerate eigenstates: $\tilde{H}_S \equiv (\tilde{H}_1^0
+ \tilde{H}_2^0) / \sqrt{2}$, $\tilde{H}_A \equiv
(\tilde{H}_1^0 - \tilde{H}_2^0) / \sqrt{2}$, $\tilde{H}^\pm $,
with eigenvalues $| m_{\tilde{H}_S}| = | m_{\tilde{H}_A} | =
| m_{\tilde{H}^\pm} | = | \mu |$: this will be sufficient for
most of the following considerations. For the discussion of chargino
and neutralino decays, it is useful to go beyond this approximation,
to see how the degeneracy is lifted. Assuming as usual $(M_2/M_1)
\simeq (3/5) \cot^2 \theta_W$, corresponding to universal gaugino
masses at some grand-unification scale, and expanding in $1/M_2$, we
find\footnote{With a slight abuse of notation, we keep the symbols
$\tilde{H}_S$, $\tilde{H}_A$ and $\tilde{H}^{\pm}$ also for the
perturbed eigenstates.}
\be
\left| m_{\tilde{H}_S} \right| = \left| \mu + \Delta_S \right| \, ,
\;\;\;
\left| m_{\tilde{H}_A} \right| = \left| \mu + \Delta_A \right| \, ,
\;\;\;
\left| m_{\tilde{H}^{\pm}} \right| = \left| \mu + \Delta_C \right| \,
,
\ee
where
\be
\label{tredelta}
\Delta_S =
( 1 - \sin 2 \beta) {4 \over 5}
{m_W^2 \over M_2} \, ,
\;\;\;\;\;
\Delta_A =
- ( 1 + \sin 2 \beta) {4 \over 5}
{m_W^2 \over M_2} \, ,
\;\;\;\;\;
\Delta_C =
- {m_W^2 \sin 2 \beta \over M_2}
\, ,
\ee
in agreement with \cite{ghappr}. We then find the mass hierarchies
\be
\begin{array}{cc}
| m_{\tilde{H}_A} | < | m_{\tilde{H}^\pm} | < | m_{\tilde{H}_S} |
&
(\mu M_2>0) \, ,
\\ & \nonumber \\
| m_{\tilde{H}_S} | < | m_{\tilde{H}^\pm} | < | m_{\tilde{H}_A} |
&
(\mu M_2<0) \, ,
\end{array}
\ee
consistent with the phenomenological request of a neutral and
weakly interacting lightest supersymmetric particle. The typical
size of the mass splittings, according to  eq.~(\ref{tredelta}),
is illustrated in fig.~\ref{sample}. Since we are not assuming
a large mixing in the stop sector, we expect radiative corrections
to the previous formulae to be negligible \cite{pierce}.

\vspace{1cm}
{\bf 2.}
In this section we present simplified analytical formulae
describing the MSSM contributions to a number of important
observables, in the special case of light $\str$ and
higgsinos, and we discuss the resulting phenomenological
constraints on the associated parameter space. Before
proceeding, we would like to state clearly the assumptions
under which the following discussion will be valid: 1)~$\str$
and higgsinos are approximate mass eigenstates, with $\stl$
and gauginos sufficiently heavy to give negligible contributions;
2)~$\tan^2 \beta < m_t/m_b$, which allows us to neglect the vertices
proportional to the bottom Yukawa coupling $h_b$, with respect to
those proportional to the top Yukawa coupling $h_t$ ($\tb <
m_t/m_b$ would be sufficient for the stop and chargino couplings,
whereas $\tan^2 \beta < m_t/m_b$ will be required by our
approximations for the charged Higgs couplings); 3)~negligible
flavour-changing effects associated with the quark-squark-gluino
and the quark-squark-neutralino vertices. Since the theoretical
expressions for the observables to be discussed below have a
strong dependence on the top quark mass, we would like to recall
here the one-loop QCD relation\footnote{In the MSSM there can be
further corrections \cite{donini} to the relation between running
and pole top quark mass, but we shall neglect them here.} between
$m_t$, the $\ov{MS}$ running mass at the top-mass scale, and the
pole mass $M_t$: $M_t = m_t [1+(4/3)\alpha_s/\pi]$. For definiteness,
we shall present our results for the input value $m_t=170$~GeV
(corresponding to $M_t \simeq 178$~GeV for $\alpha_s \simeq 0.12$),
compatible with the present Tevatron data \cite{top}.

\vspace{0.5cm}
\begin{center}
\fbox{$\Delta m_B$, $\epsilon_K$}
\end{center}

We discuss here the MSSM contributions to the $B_d^0$--$\ov{B_d^0}$
mass difference $\Delta m_{B_d}$ and to the CP-violation parameter
of the $K^0$-$\ov{K^0}$ system $\epsilon_K$, and the constraints
on the model parameters coming from the experimentally measured
values of $\Delta m_{B_d}$ and $\epsilon_K$.

For our purposes, a convenient way of parametrizing the
$B_d^0$--$\ov{B_d^0}$ mass difference is \cite{buras}:
\be
\Delta m_{B_d} = \eta_{B_d} \cdot {4 \over 3}
f_{B_d}^2 B_{B_d} \cdot m_{B_d} \cdot
\left( \alpha_W \over 4 m_W \right)^2
\cdot \left| K_{tb} K^*_{td} \right|^2
\cdot x_{tW} \cdot | \Delta | \, ,
\ee
where $\eta_{B_d} \simeq 0.55$ is a QCD correction factor;
$f_{B_d}$ is the $B_d$ decay constant and $B_{B_d}$ the vacuum
saturation parameter; $\alpha_W = g^2 / (4 \pi)$, $K$
is the Kobayashi-Maskawa matrix, $x_{tW} = m_t^2/m_W^2$.
The quantity $\Delta$ contains the
dependence on the MSSM parameters and can be decomposed as
\be
\label{dm}
\Delta = \Delta_W + \Delta_H + \tilde{\Delta} \, .
\ee
In eq.~(\ref{dm}), $\Delta_W$ denotes the Standard Model
contribution, associated with the box diagrams involving
the top quark and the $W$ boson:
\be
\Delta_W =   A ( x_{tW} ) \, ,
\ee
where the explicit expression of the function $A(x)$ is
given in the appendix. $\Delta_H$ denotes the additional
contributions from the box diagrams involving the physical
charged Higgs boson of the MSSM \cite{abbott}:
\be
\Delta_H = \cot^4 \beta \, x_{tH} {1 \over 4} G(x_{tH})
+  2 \cot^2 \beta \, x_{tW} \left[ F \, '(x_{tW}, x_{HW})
+ {1 \over 4} G \, ' (x_{tW} , x_{HW}) \right] \, ,
\ee
where $x_{tH}=m_t^2/m_{H^\pm}^2$, $x_{HW}=m_{H^\pm}^2/m_W^2$,
$\tb = v_2/v_1$, and the functions $G(x)$, $F \, ' (x,y)$
and $G \, ' (x,y)$  are given in the appendix. $\tilde{\Delta}$
denotes the contribution due to box diagrams with $R$-odd
supersymmetric particles on the internal lines. Under our
simplifying assumptions, we can take into account only the
box diagram involving the $\str$ and the charged higgsino.
The general result of \cite{bbmr} then becomes
\be
\label{dtilda}
\tilde{\Delta} =  { x_{t \tilde{\chi}} \over 4
\sin^4 \beta} G ( x_{\tilde{t} \tilde{\chi}},x_{\tilde{t}
\tilde{\chi}}) \, ,
\ee
where $x_{t \tilde{\chi}} = m_t^2 / m_{\tilde{\chi}}^2$,
$x_{\tilde{t} \tilde{\chi}} = m_{\tilde{t}}^2 /
m_{\tilde{\chi}}^2$, and $m_{\tilde{t}}$ ($m_{\tilde{\chi}}$)
is the $\str$ ($\tilde{H}^{\pm}$) mass.

Moving to the $K^0$--$\ov{K^0}$ system, the absolute value of the
parameter $\epsilon_K$ is well approximated by the expression
\cite{buras}:
\be
\vert \epsilon_K\vert=
{2 \over 3} f_{K}^2 B_{K} \cdot
\frac{m_K}{\sqrt{2} \Delta m_K} \cdot
\left( \alpha_W \over 4 m_W \right)^2
\cdot x_{cW} \cdot
| \Omega | \, ,
\label{epsk}
\ee
where $f_K$ is the $K$ decay constant, $B_K$ is the vacuum
saturation parameter, $\Delta m_K$ is the experimental
$K^0_L$--$K^0_S$ mass difference. The quantity $\Omega$,
carrying the dependence on the mixing angles and the MSSM
parameters, is given by \cite{branco}:
\be
\Omega=\eta_{cc}~ {\rm Im} (K_{cs} K_{cd}^*)^2  +
2 \eta_{ct}~ {\rm Im} (K_{cs} K_{cd}^* K_{ts} K_{td}^*)^2~
[ B(x_{tW})- \log x_{cW} ] +
\eta_{tt}~ {\rm Im} (K_{ts} K_{td}^*)^2~ x_{tc}~\Delta \, ,
\label{omega}
\ee
where $\eta_{cc} \simeq 1.38$, $\eta_{ct} \simeq 0.47$ and
$\eta_{tt} \simeq 0.57$ are QCD correction factors; $x_{cW}
= m_c^2 / m_W^2$, $x_{tc} = m_t^2/m_c^2$; the function
$B(x)$ is given in the appendix; $\Delta$ is the same as in
eq.~(\ref{dm}), and contains all the dependence on the MSSM
parameters. In principle, there are additional contributions due
to charged Higgs exchange besides those appearing in $\Delta$.
However, for $\tb \simgt 1$ they are much smaller than the
standard contribution \cite{branco}, hence they have been
neglected\footnote{In the fit to be described below, we have
explicitly checked that the inclusion of such contributions does
not modify the results appreciably.}.

We have studied the dependence of $\Delta$ on the parameters $(m_H,
\tb)$, characterizing the Higgs sector, and $(\mch,\mst)$,
characterizing the chargino-stop sector within our simplifying
assumptions (similar studies were performed in
\cite{branco,choko,coutkon}). It was already
noticed in \cite{branco} that the interference between the three
contributions
in eq.~(\ref{dm}) is always constructive, so that in general
$\Delta_{MSSM}
> \Delta_{SM}$. Besides the obvious symmetry due to the fact that
$G(1/x)=xG(x)$, in the region of parameters of present
phenomenological interest $\tilde{\Delta}$ is almost completely
controlled by $m_{ave} \equiv (\mst+\mch)/2$,
with negligible dependence on $\mst-\mch$. Given the fact that in
the MSSM, taking into account the present experimental bounds
on the neutral Higgs bosons, $m_H \simgt 100 \gev$, for stops and
charginos in the mass range accessible to LEP2, $\tilde{\Delta}$
dominates over $\Delta_H$. Moreover, due to the additional
enhancement
factor $x_{t \tilde{\chi}}$, $\tilde{\Delta}$ represents the
potentially largest contribution to $\Delta m_{B_d}$, and gives
rise to a strong dependence on $\tb$ near $\tb = 1$, due to the
$1/\sin^4 \beta$ factor in eq.~(\ref{dtilda}). Some quantitative
information is given in fig.~2, which displays contours of the
ratio
\be
\label{rdelta}
R_{\Delta} \equiv {\Delta \over \Delta_W}
\ee
in the plane $(\tb,m_{ave})$, for $m_H = 100$~GeV
(higher values of $m_H$ do not displace significantly the contours,
and we have taken for definiteness $\mst=\mch$). As can be seen,
for values of $\tb$ close to 1 and light stop and chargino,
one can obtain $R_{\Delta} \gg 1$. However, a lower limit of $\tb
\simgt 1.5$ can be obtained by requiring that the top Yukawa coupling
remain perturbative up to $M_{GUT} \sim 10^{16}$ GeV. One could
also argue that charginos lighter than 65 GeV would have been
copiously produced in the recent run of LEP~1.5, whilst no candidate
events have been reported by the standard chargino
searches \cite{lep15}. However, the reader should keep in mind that
no mass bound stronger than the LEP1 limit can be established yet
if the chargino--neutralino mass difference is sufficiently small
(a likely possibility in our approximations), or if charginos have
$R$-parity violating decays with final states consisting of jets and
no missing energy, or if the chargino production cross-section is
suppressed by the destructive interference between the
$(\gamma,Z)$-exchange and the $\tilde{\nu}_e$-exchange diagrams.
For these reasons, we think that in our analysis we can safely
consider chargino masses as low as 50 GeV or so. Thus, values of
$R_{\Delta}$ as large as about 5 can still be obtained: we shall
see in a moment how this compares with experimental data.

We now discuss the constraints coming from the measured values of
$\Delta m_{B_d}$ and $\epsilon_K$. The dependence on the MSSM
parameters is contained in the quantity $\Delta$ of eq.~(\ref{dm}),
so it would be desirable to obtain from the experimental data a
bound on $\Delta$. On the other hand, this requires some knowledge
of the parameters characterizing the mixing matrix $K$. Notice that
we cannot rely upon the SM fit to the matrix $K$, since among the
experimental quantities entering this fit there are precisely $\Delta
m_{B_d}$ and $\epsilon_K$, whose description now differs from the SM
one.

We adopt here the Wolfenstein parametrization of the mixing matrix
$K$:
\be
K=\left(
\begin{array}{ccc}
1-\dd\frac{\lambda^2}{2} & \lambda & A \lambda^3 (\rho-i\eta)\\
-\lambda & 1-\dd\frac{\lambda^2}{2} & A
\lambda^2\\
A \lambda^3 (1-\rho-i\eta) & - A \lambda^2 & 1
\end{array}
\right)
\ee
The four experimental quantities used to constrain $A$, $\rho$ and
$\eta$ are:
\begin{itemize}
\item
The direct measure of the matrix element $\vert K_{cb}\vert$,
from the semileptonic decay of the $B$ meson \cite{neubert}:
\be
\vert K_{cb}\vert=0.039\pm 0.002 \, .
\ee
This fixes the $A$ parameter, and is not affected by the MSSM
in any significant way.
\item
The direct measure of the ratio $\vert K_{ub}/ K_{cb}\vert$
from the semileptonic charmless transitions of the $B$ meson
\cite{buras}:
\be
\vert K_{ub}/ K_{cb}\vert=0.08\pm 0.02 \, .
\ee
This constrains the combination $\sqrt{\rho^2+\eta^2}$,
independently of the MSSM parameters.
\item
The $B^0$--$\ov{B^0}$ mass difference \cite{slwu}:
\be
\Delta m_{B_d}=(3.01 \pm 0.13) \times 10^{-13}~{\rm GeV} \, .
\ee
This constrains the combination $A^2 [(1-\rho)^2+\eta^2]$, as in the
SM. However, it depends on the MSSM parameters through $\Delta$.
\item
The parameter $\epsilon_K$ of CP violation \cite{buras}:
\be
\vert \epsilon_K \vert= (2.26\pm 0.02) \times 10^{-3} \, .
\ee
Here one tests an independent combination of $(A,\rho,\eta)$, which
depends on the value of $\Delta$ in the MSSM.
\end{itemize}

To derive the desired bound on $\Delta$, we have performed a fit to
these data suitable for the MSSM, i.e. keeping $A$, $\rho$, $\eta$
and the $\Delta$ as independent variables. The results of the fit
are sensitive to the input values of the parameters $f_{B_d}^2
B_{B_d}$ and $B_K$. We have repeated the fit for various allowed
values of $f_{B_d}^2 B_{B_d}$ and $B_K$, to estimate the effect
of the corresponding theoretical uncertainties. We have checked
that, by fixing $\Delta$ to its SM value, $\Delta = \Delta_W =
0.551$ for $m_t = 170~{\rm GeV}$, we recover the results for
$(A,\rho,\eta)$ of the SM fit \cite{buras}.

\begin{table}[hbtp]
\begin{center}
\begin{tabular}{|c|c|c|c|}
\hline
$\sqrt{f_{B_d}^2 B_{B_d}}~({\rm MeV})$ & $\rho$ & $\eta$ & $\Delta$\\
\hline
160 & $-0.19\pm 0.14$ &
      $ 0.31\pm 0.04$ & $0.55\pm 0.15$\\
\hline
160 & $+0.30^{+0.12}_{-0.61}$ &
      $0.21^{+0.14}_{-0.04}$ & $1.52^{+0.70}_{-1.10}$ \\
\hline
180 & $-0.08\pm 0.23$ &
      $ 0.35\pm 0.05$ & $0.50\pm 0.22$\\
\hline
180 & $+0.23\pm 0.22$ &
      $0.28\pm 0.08$ & $0.97\pm 0.52$ \\
\hline
200 & $+0.09\pm 0.41$ &
      $ 0.37\pm 0.11$ & $0.54\pm 0.46$\\
\hline
220 & $+0.10^{+0.21}_{-0.22}$ &
      $0.41^{+0.06}_{-0.07}$ & $0.43^{+0.27}_{-0.15}$ \\
\hline
240 & $+0.10^{+0.18}_{-0.19}$ &
      $0.45^{+0.06}_{-0.07}$ & $0.34^{+0.17}_{-0.10}$ \\
\hline
\end{tabular}
\end{center}
\caption{Results of the fit for $m_t=170~{\rm GeV}$,
$B_K=0.75$ and for different values of $f_{B_d}^2 B_{B_d}$.
The fitted value of $A$ ranges from 0.80 to 0.83, with an
uncertainty of 0.04.}
\end{table}

In table~1 we show our results for several choices of
$f_{B_d}^2 B_{B_d}$ and for $B_K=0.75$.
In all cases the parameter $A$, basically determined by $\vert
K_{cb}\vert$, essentially coincides with its SM determination.
For relatively low values of $f_{B_d}^2  B_{B_d}$, small values of
$\eta$ are preferred and the $\chi^2$ function has two minima: this
is due to the constraint coming from $\vert K_{ub}/ K_{cb} \vert$,
which is sensitive to $\sqrt{\rho^2 + \eta^2}$ and, for positive
$\eta$, has a twofold ambiguity in $\rho$. For $\sqrt{f_{B_d}^2
B_{B_d}} \simlt 190$~MeV, the negative $\rho$
solution is the one favoured by the SM and leads to a central value
for $\Delta$ which is very close to the SM one. On the contrary,
the positive $\rho$ solution, which is absent in the SM for the
current choice of parameters, corresponds to a higher $\Delta$.
For relatively high values of $f_{B_d}^2 B_{B_d}$, the $\chi^2$
function has a unique minimum, $\rho$ is very close to zero and
$\Delta$ is close to its SM value. It is clear that, qualitatively,
large values of $\Delta$ can be allowed only for small $f_{B_d}^2
B_{B_d}$.

In table~2 we show the influence of the $B_K$ parameter.
The highest values of $\Delta$ are obtained when $B_K=0.9$. This can
be qualitatively understood as follows: a large $\Delta$ is
compatible with the measured $\Delta m_{B_d}$ only when $\eta$ is
close to zero. On the other hand, since $\vert \epsilon_K\vert$ is
proportional to $\eta$, the smallness of $\eta$ should be compensated
by $B_K$, which is then required to be large.
\begin{table}[hbtp]
\begin{center}
\begin{tabular}{|c|c|c|c|c|c|}
\hline
$\sqrt{f_{B_d}^2 B_{B_d}}~({\rm MeV})$ & 160 & 180 & 200 & 220 & 240
\\
\hline
$B_K=0.6$ & $1.12^{+0.73}_{-0.65}$ & $0.67\pm 0.39$ &
            $0.51^{+0.27}_{-0.16}$ & $0.39^{+0.17}_{-0.11}$ & \\
\hline
$B_K=0.9$ & $1.69^{+0.72}_{-0.50}$ & $1.20^{+0.54}_{-0.86}$ &
            $0.82^{+0.43}_{-0.39}$ & $0.46\pm 1.54$ &
            $0.36^{+0.26}_{-0.13}$ \\
\hline
\end{tabular}
\end{center}
\caption{Central values and 1$\sigma$ errors for $\Delta$, for
different
choices of $f_{B_d}^2 B_{B_d}$ and $B_K$. When two minima are
present, only the largest central value for $\Delta$ is quoted.}
\end{table}

It is not straightforward to translate the above results into a
single definite bound on $\Delta$, or, equivalently, on $R_{\Delta}
= \Delta / \Delta_W \simeq 1.8 \Delta$. Values of $R_{\Delta}$ as
large as 5 (see fig.~2) are clearly disfavoured, but cannot be
rigorously excluded  if one keeps in mind the theoretical
uncertainties on the parameters $f_{B_d}^2 B_{B_d}$ and $B_K$.
For the moment, we can only state that small $\tan\beta$ and very
light chargino and stop require small $f_{B_d}^2 B_{B_d}$ and
large $B_K$, and imply small $\eta$ and positive and large $\rho$.

The expression of $\Delta m_{B_s}$ can be trivially obtained from
eq.~(\ref{sample}) by making everywhere the replacement $d
\rightarrow s$.  The present $95\%$ CL limit \cite{slwu}, $\Delta
m_{B_s} \ge 4.0 \times 10^{-12}~{\rm GeV}$, does not provide
additional
constraints on $\Delta$. One obtains:
\be
x_{tW} \Delta \ge 1.14 \left( \frac{230}{f_{B_s}
\sqrt{B_{B_s}}~({\rm MeV})}\right)^2 \left(
\frac{0.8}{A} \right)^2 \, .
\ee
In the SM, for $m_t=170~{\rm GeV}$ one has $x_{tW} \Delta_W=2.47$.
Since $\Delta\ge\Delta_W$, the previous limit is always respected
in the MSSM, for all values of the parameters. On the other hand,
we can derive some information on $\Delta m_{B_s}$ in the MSSM
from the relation:
\be
\frac{\Delta m_{B_s}}{\Delta m_{B_d}}=
\frac{m_{B_s}}{m_{B_d}}\cdot \xi_s^2 \cdot
\frac{\eta_{B_s}}{\eta_{B_d}}
\left\vert\frac{K_{ts}}{K_{td}}\right\vert^2 \, ,
\label{sdratio}
\ee
where
$\xi_s=[ f_{B_s} \sqrt{B_{B_s}}] / [f_{B_d} \sqrt{B_{B_d}}]$.
One expects $\xi_s = 1.16 \pm 0.1$ \cite{xi} and $\eta_{B_s}
= \eta_{B_d}$. Then, from eq.~(\ref{sdratio}) one obtains:
\be
\Delta m_{B_s}\simeq (28\pm 5)\cdot \frac{\Delta m_{B_d}}
{\left[(1-\rho)^2+\eta^2\right]} \, .
\ee
This relation is valid both in the SM and in the limit of the
MSSM considered here. However, the high value of $\Delta$ which
could be obtained in the MSSM for small $\tb$ and light chargino
and stop, would imply a value for the combination $(1-\rho)^2
+\eta^2$ smaller than in the SM, as can be seen from table~1.
On this basis, we conclude that the value expected for $\Delta
m_{B_s}$ in the MSSM, when $\tb$ is small and stop and chargino
are both light, is always higher than the one foreseen in the SM.
However, in view of the existing uncertainties on $\rho$ and $\eta$,
a more precise estimate of $\Delta m_{B_s}$ in the MSSM is not yet
possible.

To conclude this section, we would like to comment on the MSSM
effects on the ratio $\epsilon'/\epsilon$. These have been
analysed, at leading order in QCD and QED, in ref.~\cite{gab}.
We recall that, on the experimental side, there are two
independent results for ${\rm Re}~ (\epsilon'/\epsilon)$:
\be
\begin{array}{ccc}
{\rm Re \,} (\epsilon' / \epsilon) &=&23\pm 6.5
\times 10^{-4}~~~~~~ {\rm NA31~~~\cite{NA31}\, ,} \\
{\rm Re \,} ( \epsilon' / \epsilon) &=&7.4\pm 6.0
\times 10^{-4}~~~~~~{\rm E731~~~\cite{E731}\, .}
\end{array}
\ee
The SM value of ${\rm Re \,} \epsilon'/\epsilon$ is typically of
order $10^{-4}$ for $m_t=150$--$190~{\rm GeV}$, decreases for
increasing $m_t$, and vanishes for $m_t=200$--$220~{\rm GeV}$.
In the MSSM, it is possible to enhance the SM prediction by at most
$40\%$ for $m_t \simeq 170~{\rm GeV}$ and up to $60\%$ for $m_t
\simeq 190~{\rm GeV}$. The enhancement is attained for chargino
and stop masses close to the present LEP limit, with the other
squarks and the charged Higgs much heavier. This modest enhancement
cannot explain the large central value of ${\rm Re \,} \epsilon'/
\epsilon$ suggested by the NA31 experiment and, on the other hand,
is perfectly compatible with the data of the E731 collaboration.
A reduction of ${\rm Re \,} \epsilon'/\epsilon$ with respect to the
SM value is also achievable in the MSSM. This requires a light
charged Higgs and light charginos and stops. Part of the effect is
due to the fact that ${\rm Re \,} \epsilon'/\epsilon$ is proportional
to $\eta$, which, as discussed above, can be considerably smaller
than in the SM. In this case a vanishing or even negative value of
${\rm Re \,} \epsilon'/\epsilon$ can be obtained for
$m_t=150$--$190~{\rm GeV}$. This depletion, which potentially
represents the most conspicuous effect of minimal supersymmetry,
is however very difficult to test, due to the insufficient
experimental sensitivity. In conclusion, the present data on
$\epsilon'/\epsilon$ do not provide any additional constraint on
the MSSM parameter space.

\vspace{0.5cm}
\begin{center}
\fbox{$b \to s \gamma$}
\end{center}
The recent CLEO result \cite{cleo} on the inclusive $B \to X_s
\gamma$ decay, $BR(B \to X_s \gamma) = (2.32 \pm 0.67) \times
10^{-4}$, agrees with the SM predictions based on the partonic
process $b \to s \gamma$ (for a review and references, see e.g.
\cite{ricciardi}), and at the same time constrains possible
extensions of the SM, in particular the MSSM. A very conservative
estimate \cite{ali} gives $BR(B \to X_s \gamma)_{SM} = (2.55 \pm
1.28) \times 10^{-4}$, whereas other authors \cite{altribsg}
quote similar central values but smaller errors, at the level
of $30 \%$. Under our simplifying assumptions, the $b \to s \gamma$
amplitude
at a scale ${\cal O}(M_W)$ receives additional contributions from
top and charged Higgs (stop and chargino) exchange, which interfere
constructively (destructively) with the SM contributions, dominated
by top and $W$ exchange. The amplitude at a scale ${\cal O} (m_b)$
gets both multiplicatively and additively renormalized by QCD
corrections. The latter effect is mainly due to the mixing between
the magnetic operator ($O_7$) and a four-quark operator ($O_2$).
We will express our results in terms of the ratio
\be
\label{rgamma}
R_{\gamma} \equiv
\frac{Br (B \to X_s \gamma)_{MSSM}}{Br(B \to X_s \gamma)_{SM}}
\, ,
\ee
which we identify with the corresponding ratio of
$b \to s \gamma$ squared amplitudes. Then we estimate:
\be
R_{\gamma} \simeq
\left[ \frac{ C (A_W + A_H + {\tilde A}) + D }
 { C~ A_W + D } \right]^2 \, ,
\ee
where \cite{bbmr} $C \simeq 0.66$, $D \simeq 0.35$,
\be
A_W = x_{tW} ( 2 F_1(x_{tW}) + 3 F_2 (x_{tW}) ) \, ,
\ee
\be
A_H =
x_{tH} \left\{ \cot^2 \beta \left[ {2 \over 3} F_1(x_{tH}) + F_2
(x_{tH}) \right] + \left[ {2 \over 3} F_3(x_{tH}) + F_4 (x_{tH})
\right] \right\} \, ,
\ee
\be
\tilde{A} = -
{x_{t \tilde{t}} \over \sin^2 \beta} \left[ F_1(x_{\tilde{\chi}
\tilde{t}}) + {2 \over 3} F_2 (x_{\tilde{\chi} \tilde{t}}) \right]
\, .
\ee

Similarly to the previously discussed $R_{\Delta}$, we have
found that $R_{\gamma}$ depends on $\mst$ and $\mch$
essentially through their sum (not their difference).
Then we can focus as before on the variable $m_{ave}=
(\mst+\mch)/2$. Figs.~3 and 4 show contour lines of
$R_{\gamma}$, in the ($m_{ave},m_H$) plane for $\tb=1.5,5$
and in the ($m_{ave},\tb$) plane for $m_H=100,500 \gev$,
respectively, taking for definiteness $\mst=\mch$. The
contour $R_{\gamma}=1$ corresponds to the situations in which
the `extra' contributions $A_H$ and ${\tilde A}$
cancel against each other (the possibility of these
cancellations was emphasized in ref.~\cite{bargiu}), so that
the SM result is recovered. Since the comparison between
theory and experiment is dominated by the theoretical error,
the allowed region can be estimated conservatively to be $0.5
\simlt R_{\gamma} \simlt 1.5$, or slightly less conservatively
$0.7 \simlt R_{\gamma} \simlt 1.3$. In fig.~3, one can notice the
strong positive correlation between $m_{ave}$ and $m_H$. In other
words, light charged Higgs and heavy stop and chargino would give
too large a value for $BR(B \to X_s \gamma)$, whereas heavy charged
Higgs and light stop and chargino would give too small a value.
In addition, fig.~4 shows a moderate dependence on $\tb$ in the
range $1 \simlt \tb \simlt 2$.

\vspace{0.5cm}
\begin{center}
\fbox{$R_b$}
\end{center}
The calculation of $R_b \equiv \Gamma(Z \to b \ov{b})/\Gamma
(Z \to {\rm hadrons})$ in the MSSM was performed in~\cite{boulfin}.
Specializing those results to the limiting case under discussion,
we can write
\be
R_b = (R_b)_{SM} \left[ 1 + 0.78 \times
{\alpha_W \over 2 \pi} {v_L  \over v_L^2 + v_R^2}
(F_H + \tilde{F}) \right] \, ,
\ee
where, for the input values $M_t=180 \pm 12 \gev$ and $\alpha_S
(m_Z) = 0.125 \pm 0.007$,
\be
(R_b)_{SM} = 0.2156 \pm 0.0005 \, ,
\ee
and
\be
v_L = - {1 \over 2} + {1 \over 3} \sin^2 \theta_W \, ,
\;\;\;\;\;
v_R = {1 \over 3} \sin^2 \theta_W \, .
\ee
$F_H$ and $\tilde{F}$ are associated with top-Higgs and
stop-higgsino loops, respectively, and read
\be
\begin{array}{c}
F_H = \left\{ b_1(m_H,m_t) v_L - c_0 (m_t,m_H) v_L^{(H)}
+ m_t^2 c_2(m_H,m_t) v_L^{(t)} \right.
\\
+ \left.
\left[ m_Z^2 c_6(m_H,m_t) - {1 \over 2}
- c_0(m_H,m_t) \right] v_R^{(t)} \right\}
\lambda_H^2 \, ,
\end{array}
\ee
\be
\begin{array}{c}
\tilde{F} = \left\{ b_1(\mst,\mch) v_L
- c_0 (\mch,\mst) v_R^{(t)} \right.
\\
\left. + \left[ m_Z^2 c_6(\mst,\mch) - {1 \over 2}
- c_0(\mst,\mch) + \mch^2 c_2(\mst,\mch)
\right] v_L^{(H)} \right\} \tilde{\lambda}^2 \, ,
\end{array}
\ee
where the functions $b_1$, $c_0$, $c_2$ and $c_6$ are
given in the appendix, and
\be
\lambda_H = {m_t \over \sqrt{2} m_W \tb} \, ,
\;\;\;\;\;
\tilde{\lambda} = {m_t \over \sqrt{2} m_W \sin \beta} \, ,
\ee
\be
v^{(t)}_L = {1 \over 2} - {2 \over 3} \sin^2 \theta_W \, ,
\;\;\;\;\;
v^{(t)}_R = - {2 \over 3} \sin^2 \theta_W \, ,
\;\;\;\;\;
v^{(H)}_L = - {1 \over 2} + \sin^2 \theta_W \, .
\ee

A quantitative estimate of the possible effects is given in
fig.~5, which shows contours of $R_b$ in the $(\mch,\mst)$
plane, for some representative values of $\tb$ and $m_H$.
One can see that, in our limiting case, values of $R_b$
as high as $0.218$ can be reached, for $\str$ and higgsinos very
close to 50~GeV. Notice also that the dependence of $R_b$
on $m_{\tilde{\chi}}$ is stronger than the dependence on
$m_{\tilde{t}}$, which makes the higgsino mass $\mu$ the most
relevant parameter. The dependences on $\tb$ and on $m_H$
are not very strong, and the effect is maximal for
$m_{\tilde{\chi}}$,
$m_{\tilde{t}}$ as close as possible to their experimental limits,
low $\tb$ (maximal top Yukawa coupling) and high $m_H$ (minimal
negative interference with the charged Higgs loops).

In the past, it was suggested \cite{fits} that an improved fit
to $\alpha_S(m_Z)$
and to $R_b$ could be obtained by allowing for some new physics that
enhances $R_b$ with respect to its SM prediction. The most recent
experimental data \cite{ewwg} give $R_b=0.2219 \pm 0.0017$, with
a strong positive correlation with $R_c = 0.1540 \pm 0.0074$, which
also significantly deviates from its SM prediction, $(R_c)_{SM} =
0.1724 \pm 0.0003$. If one fixes $R_c$ to its SM value, the fit to
the LEP data gives $R_b=0.2205 \pm 0.0016$. Even in the last, most
favourable case, our limiting case of the MSSM cannot produce $R_b$
closer than $1.5 \sigma$ to its experimental value. Slightly better
agreement can be obtained for very large values of $\tb$ and $A^0$
as light as allowed by the present experimental limits, but this
case cannot be quantitatively studied within the present
approximations.

\vspace{0.5cm}
\begin{center}
\fbox{$t \to \tilde{t} \tilde{H}_{S},
\; \tilde{t} \tilde{H}_{A}, \; b H^+$}
\end{center}
In the presence of sufficiently light charged Higgs boson, stop and
higgsinos, new decay modes are kinematically accessible in the top
quark decays, in addition to the standard mode $t \to b W^+$:
assuming heavy sbottom squarks, they are $t \to \tilde{t}
\tilde{H}_{S} , \; \tilde{t} \tilde{H}_{A}, \; b H^+$. The
corresponding partial widths are reported below \cite{ridolfi}:
\be
\Gamma (t \to \tilde{t} \tilde{H}_i^0 ) =
{\sqrt{
[m_t^2 - (m_{\tilde{t}} + m_{\tilde{H}_i})^2]
[m_t^2 - (m_{\tilde{t}} - m_{\tilde{H}_i})^2]}
\over 16 \pi m_t^3} \cdot {\cal A}_i \, ,
\ee
\be
{\cal A}_i  = {g^2 m_t^2  \over 8 m_W^2 \sin^2 \beta}
( m_t^2 + m_{\tilde{H}_i}^2 -m_{\tilde{t}}^2) \, ,
\;\;\;\;\; (i=S,A) \, ;
\ee
\be
\Gamma (t \to b H^+ ) =
{\sqrt{
[m_t^2 - (m_H + m_b)^2]
[m_t^2 - (m_H - m_b)^2]}
\over 16 \pi m_t^3} \cdot {\cal A}_H \, ,
\ee
\be
{\cal A}_H  = {g^2 m_t^2 \over 4 m_W^2 \tan^2 \beta}
( m_t^2 + m_b^2 -m_H^2) \, ;
\ee
\be
\Gamma (t \to b W^+ ) =
{\sqrt{
[m_t^2 - (m_W + m_b)^2]
[m_t^2 - (m_W - m_b)^2]}
\over 16 \pi m_t^3} \cdot {\cal A}_W \, ,
\ee
\be
{\cal A}_W  = {g^2 \over 4} \left[ m_t^2 + m_b^2 - m_W^2
- {m_W^4 - (m_t^2 - m_b^2)^2 \over m_W^2} \right] \, .
\ee
With the help of fig.~6, which displays contours of $BR(t \to b
W^+)$ in the $(\mu,\mst)$ plane, for some representative values
of $\tb$ and $m_H$, we can see that deviations from the SM
prediction $BR(t \to b W^+) \simeq 1$ can be very significant,
up to $BR(t \to b W^+) \sim 0.4$. However, this cannot be
transformed easily into a constraint on the parameter space, as
attempted in \cite{mang}: first, the perturbations to our
limiting case, illustrated in fig.~1, can modify the results
for the top branching ratios, but cannot be accounted for without
introducing additional parameters such as $M_2$;
second, it is not clear how strong a lower bound the present CDF and
D0 data can provide on $BR(t \to b W^+)$: deriving such a bound
requires not only the detailed knowledge of the experimental
selection criteria, but also assumptions about the production
cross-section and the stop and higgsinos branching ratios. We do
not feel in a position to do so reliably, so we content ourselves
with displaying the contours in fig.~6.  As a reference value
for the CDF and D0 sensitivity, we can tentatively take $BR (
t \rightarrow b W^+) = 0.7$: it is then clear that the
foreseeable Tevatron bounds on exotic top decays will
significantly constrain the light $\str$-higgsino scenario,
in qualitative agreement with the conclusions of ref.~\cite{mang}.

\vspace{1cm}
{\bf 3.}
Higgsinos and $\str$ are among the most plausible light
supersymmetric particles, and a mass range for these states
within the experimental reach of LEP2 would represent an
appealing scenario. At the moment, this range is still
compatible with the precision electroweak data, and it
could even play a role in reducing the present discrepancy
concerning $R_b$. A more systematic analysis is however
required to confront this possibility with the available
experimental information, including the rich input coming
from flavour-changing transitions. The aim of this note
has been to improve the existing studies, which often focus
on a single specific process, by discussing all the relevant
constraints, albeit in a simplified setting.

Among the observables that are potentially most sensitive,
we focused on the mixing parameter $\Delta m_{B_{d}}$
and the CP--violating parameter $\epsilon_K$. One could
have expected that, given the present experimental precision
on those data, light $\str$ and higgsinos could already be
ruled out, at least for small $\tb$. Actually, due to the
theoretical uncertainties affecting $f^2_{B_d} B_{B_d}$ and
$B_K$, we have been led to a milder statement. The corner in
the MSSM parameter space with very light $\str$ and higgsino,
and small $\tan\beta$, requires small values of $f^2_{B_d} B_{B_d}$
and large ones for $B_K$, at the border of the presently allowed
theoretical ranges. Moreover, positive and large values of $\rho$
and  small values of $\eta$ are preferred, with immediate
implications on the values of the CP--violating asymmetries
in $B$ decays at future facilities.

The process that gives the most significant constraints on the
reduced MSSM parameter space, corresponding to the limiting case
of light $\str$ and higgsinos, is $b \rightarrow s \gamma$ (see
figs.~3 and 4). For example, the existence of $\str$ and
higgsinos around 60~GeV would require\footnote{These bounds
could be somewhat relaxed, however, when mixing effects in
the chargino and stop sectors are included.}
$m_H \simlt 100 \gev$ for $\tb = 1.5$, $m_H \simlt 200 \gev$
for $\tb = 5$. This has also indirect effects on the allowed values
for the neutral Higgs bosons of the MSSM. Since in our
limiting case the sum rule $m_A^2 + m_W^2 = m_H^2$ remains
valid to quite a good accuracy after the inclusion of
radiative corrections \cite{charged}, one gets corresponding
approximate upper bounds on $m_A$. The mass of the lightest
CP-even state could also be affected, since its tree-level value
depends on $(m_A,\tb)$, whilst radiative corrections \cite{neutral}
are mainly controlled by the logarithmic dependence on
$m_{\tilde{t}_1} m_{\tilde{t}_2}$, for fixed $M_t$ and $\tb$.
Under our assumptions, however, we are still free
to push $m_{\stl}$ to values sufficiently high that the
experimental bounds can be evaded.

The measurement of $R_b$ at LEP and the study of top decays at
the Tevatron cannot be transformed, for the moment, into precise
bounds on the MSSM parameter space. In the case of $R_b$, besides
the open question of the correlation with $R_c$, the size of the
typical effects of light $\str$ and higgsinos is considerably
smaller than the discrepancy between the SM prediction and the
experimental average. In the case of top decays, only the CDF
and D0 collaborations have the appropriate tools to establish
reliable bounds on exotic channels, either directly or by the
extraction of $BR(t \rightarrow b W^+)$. If, as expected, the
bound setlles around $BR(t \rightarrow b W^+) > 0.7$, then the
surviving region of the $(\mch,\mst)$ plane will allow at most
for $\Delta R_b \simlt 10^{-3}$, a rather marginal improvement
over the SM when compared with the experimental data.

Finally, we recall that the parameter space discussed in section~2
starts to get significant constraints from the direct searches for
stops and charginos, both at the Tevatron \cite{teva} and at LEP
\cite{lep15,lep}. We hope that the analysis reported in this paper
will contribute to the understanding of the interplay between
indirect and direct signals of light $\str$ and higgsinos.

\section*{Acknowledgements}
We would like to thank G.~Altarelli, M.~Carena, P.~Checchia,
M.~Ciuchini, J.~Ellis, U.~Gasparini, G.F.~Giudice, M.~Mangano,
G.~Ridolfi and C.E.M.~Wagner for useful discussions.
\newpage
\section*{Appendix}
We collect in this appendix the functions, obtained from one-loop
diagrams, appearing in the formulae given in the text.
\be
\label{adix}
A(x) = {1 \over 4 (x-1)^3} (x^3 - 12 x^2 + 15 x - 4 + 6 x^2
\log x) \, ,
\ee
\be
G(x) = {-1 + x^2 - 2 x \log x \over (x-1)^3} \, ,
\ee
\be
F \, '(x,y) =
{ -1 + x - \log x \over (x-1)^2(y-x)} +
{\displaystyle
{x \log x \over x-1}+{y \log y \over 1 - y}
\over (x-y)^2}
 \, ,
\ee
\bea
G \, '(x,y) & = &
{3 - 4 x + x^2  + 4 x \log x - 2 x^2  \log x \over
2 (x-1)^2  (y-x)}
\nonumber \\
& & \nonumber \\
& - &
{\displaystyle
{3 (y-x) \over 2} + {x^2 \log x \over x - 1} +
{y^2 \log y \over 1 - y} \over (x-y)^2 }
\, ,
\eea
\be
B(x) = \log x - \frac{3}{4} \frac{x}{x-1}\left( -1 +
\frac{x}{x-1} \log x \right)\, ,
\ee
\be
F_1(x) = {1 \over 12 (x-1)^4} (x^3 - 6 x^2 + 3 x + 2 + 6 x \log x) \,
,
\ee
\be
F_2(x) = {1 \over 12 (x-1)^4} (2 x^3 + 3 x^2 - 6 x + 1 - 6 x^2 \log
x) \, ,
\ee
\be
F_3(x) = {1 \over 2 (x-1)^3} (x^2 - 4 x + 3 + 2 \log x) \, ,
\ee
\be
F_4(x) = {1 \over 2 (x-1)^3} (x^2 -1 - 2 x \log x) \, .
\ee
\be
b_1(m_1,m_2) = - {1 \over 4} + {m_2^2 \over 2 (m_1^2 - m_2^2)}
+ {\log (m_1^2/\mu^2) \over 2} + {m_2^4 \log (m_2^2/m_1^2)
\over 2 (m_1^2 - m_2^2)^2} \, ,
\ee
\be
\begin{array}{cl}
c_0(m_1,m_2) = &
\displaystyle \int_0^1 dx \left\{
-{(m_1^2 - m_1^2 x + m_2^2 x) [1 - \log (m_1^2 - m_1^2 x
+ m_2^2 x)/\mu^2] \over m_1^2 - m_2^2 + m_Z^2 x} \right.
\\ & \displaystyle \left.
+ {(m_2^2 - m_Z^2 x + m_Z^2 x^2) [1 - \log (m_2^2 - m_Z^2
x + m_Z^2 x^2)/\mu^2] \over m_1^2 - m_2^2 + m_Z^2 x} \right\}
\, ,
\end{array}
\ee
\be
c_2(m_1,m_2) = \int_0^1 dx {\log [(m_1^2 - m_1^2 x +
m_2^2 x)/(m_2^2 - m_Z^2 x + m_Z^2 x^2)] \over m_1^2 - m_2^2
+ m_Z^2 x} \, ,
\ee
\be
c_6(m_1,m_2) = \int_0^1 dx {x \log [ (m_1^2 - m_1^2
x + m_2^2 x) / (m_2^2 - m_Z^2 x + m_Z^2 x^2) ] \over m_1^2
- m_2^2 + m_Z^2 x} \, .
\ee
\newpage
\newpage
\vfill{
\begin{figure}[h]
\vspace{-0.1cm}
\centerline{
\epsfig{figure=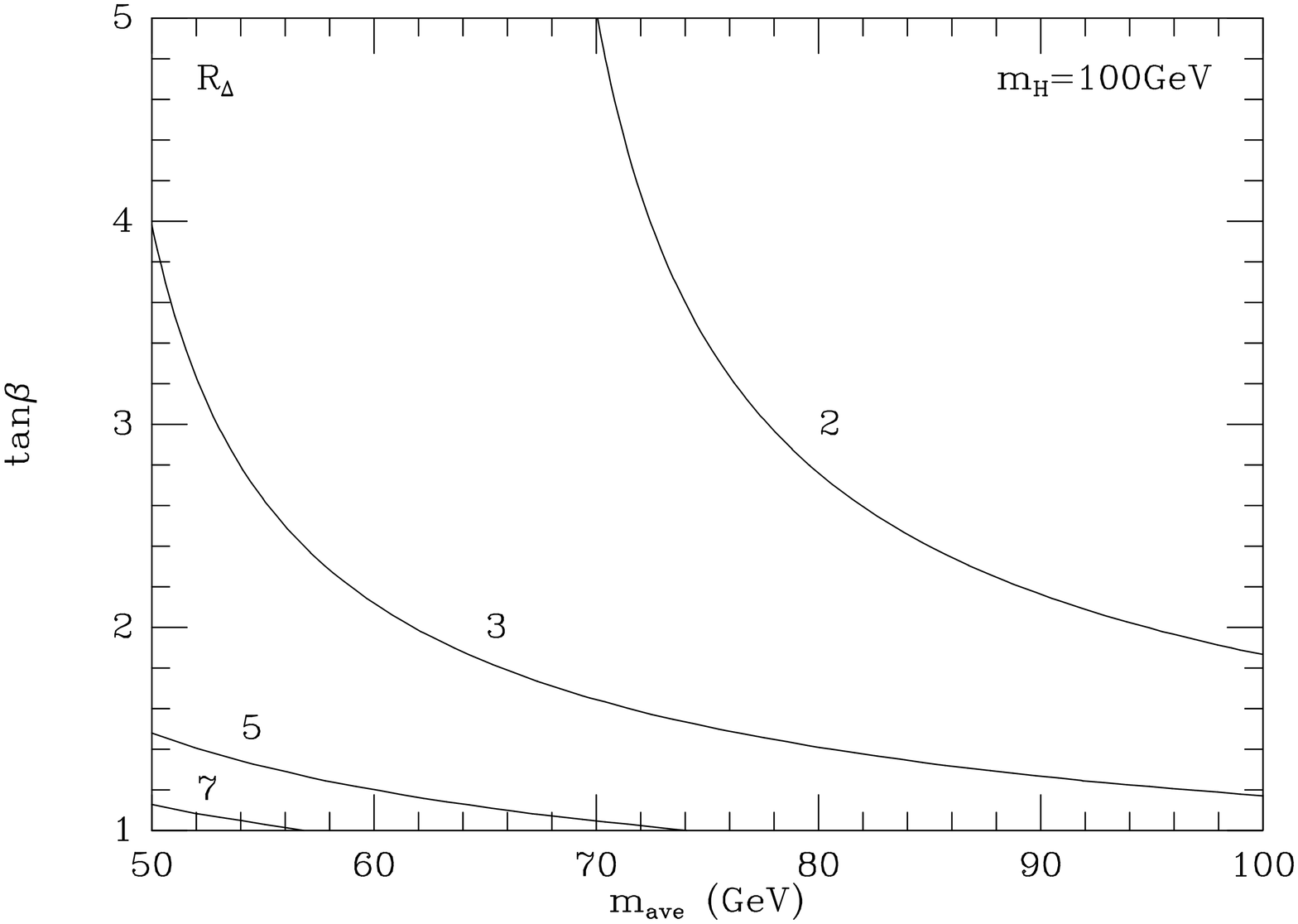,height=9cm,angle=-90}
}
\vspace{-0.1cm}
\caption{\it Contours of $R_{\Delta}$ in the
($m_{ave},\tb$) plane, for $m_H = 100$ GeV.}
\label{fig2}
\end{figure}
}
\newpage
\begin{figure}[h]
\centerline{
\epsfig{figure=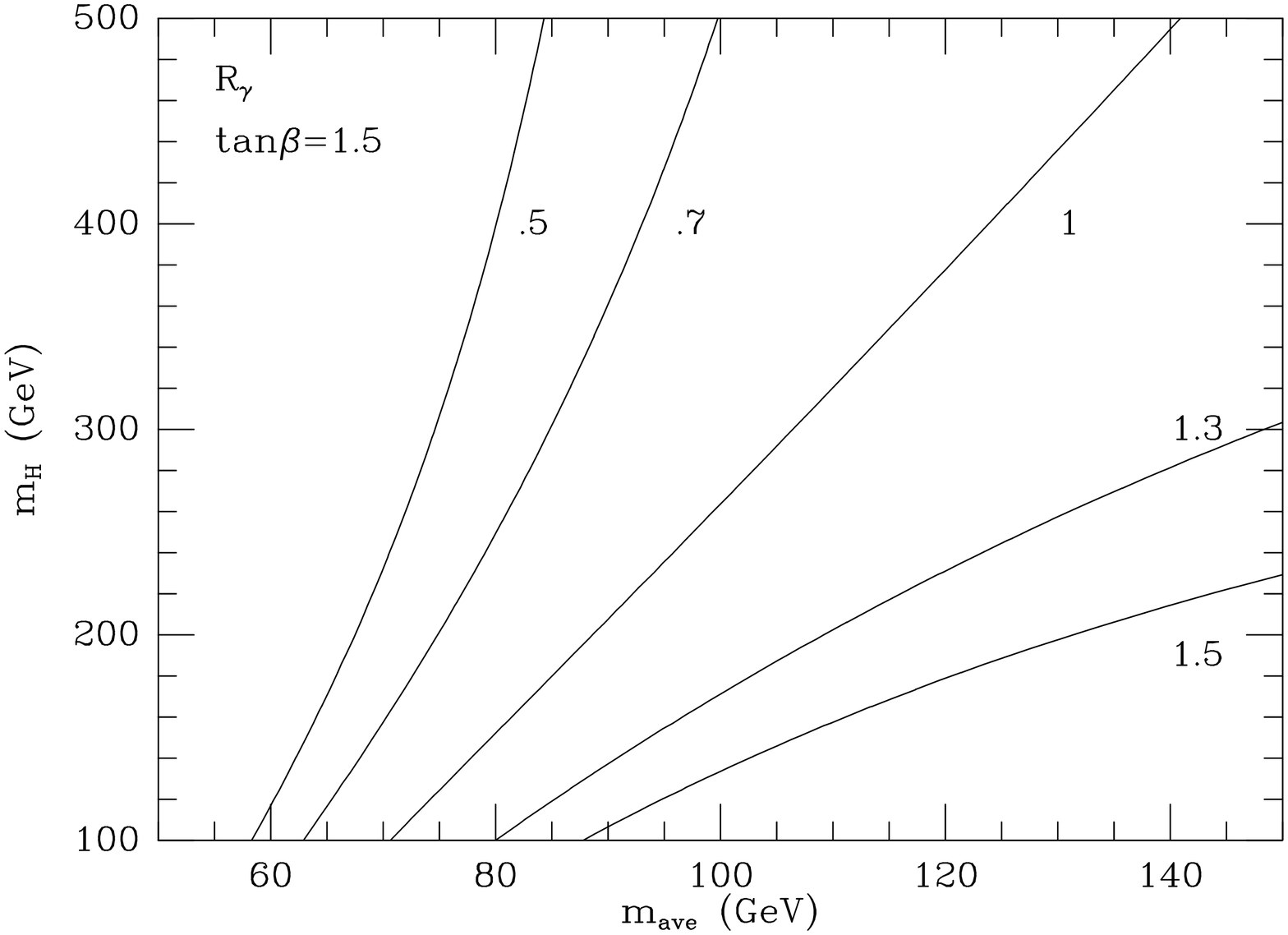,height=9cm,angle=-90}}
\vspace{0.1cm}
\centerline{
\epsfig{figure=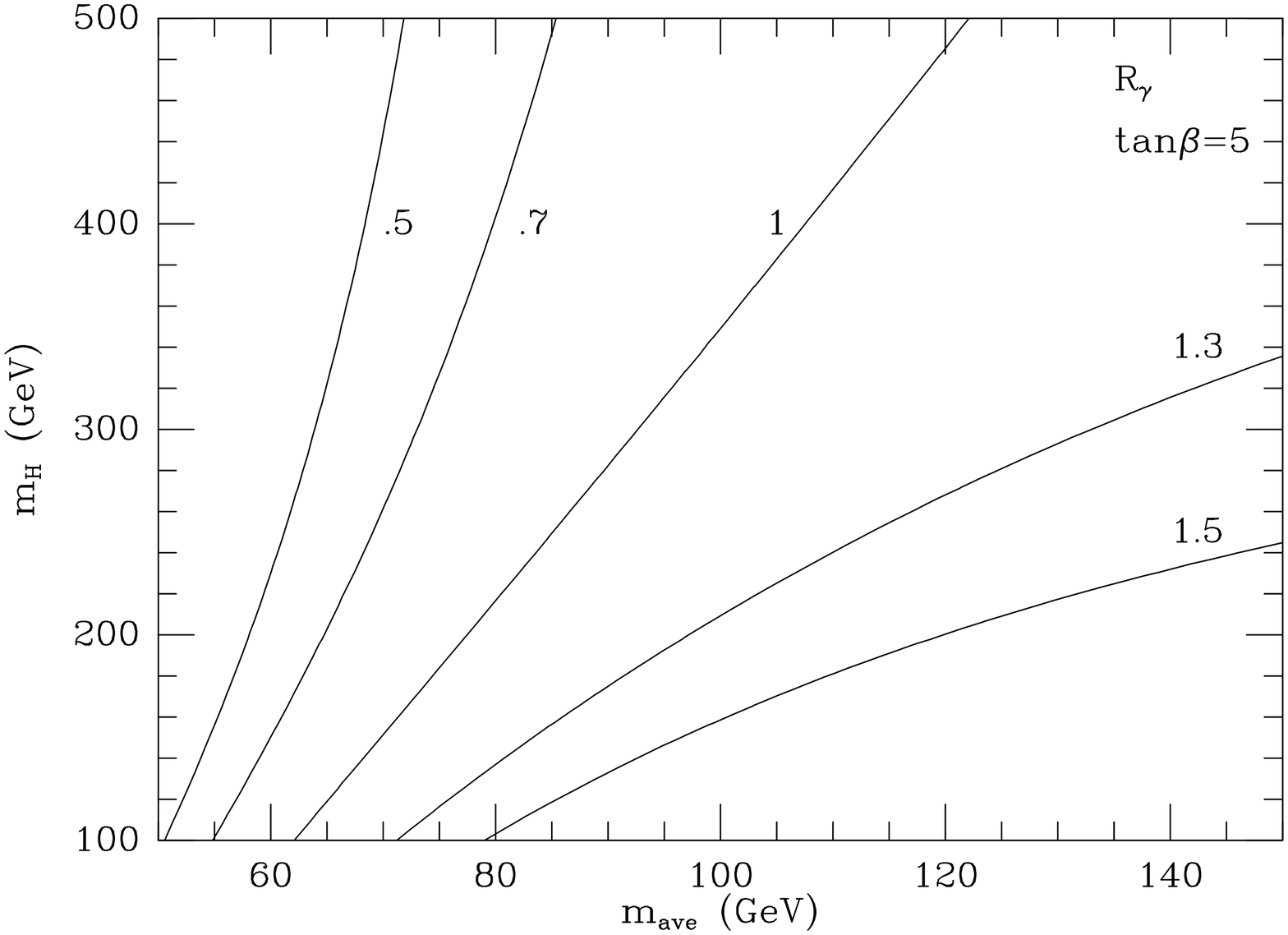,height=9cm,angle=-90}}
\vspace{0.1cm}
\caption{{\it Contours of $R_{\gamma}$ in the
$(m_{ave},m_H)$ plane, for $\tb = 1.5,5$.}}
\label{fig3uno}
\end{figure}
\newpage
\begin{figure}[h]
\centerline{
\epsfig{figure=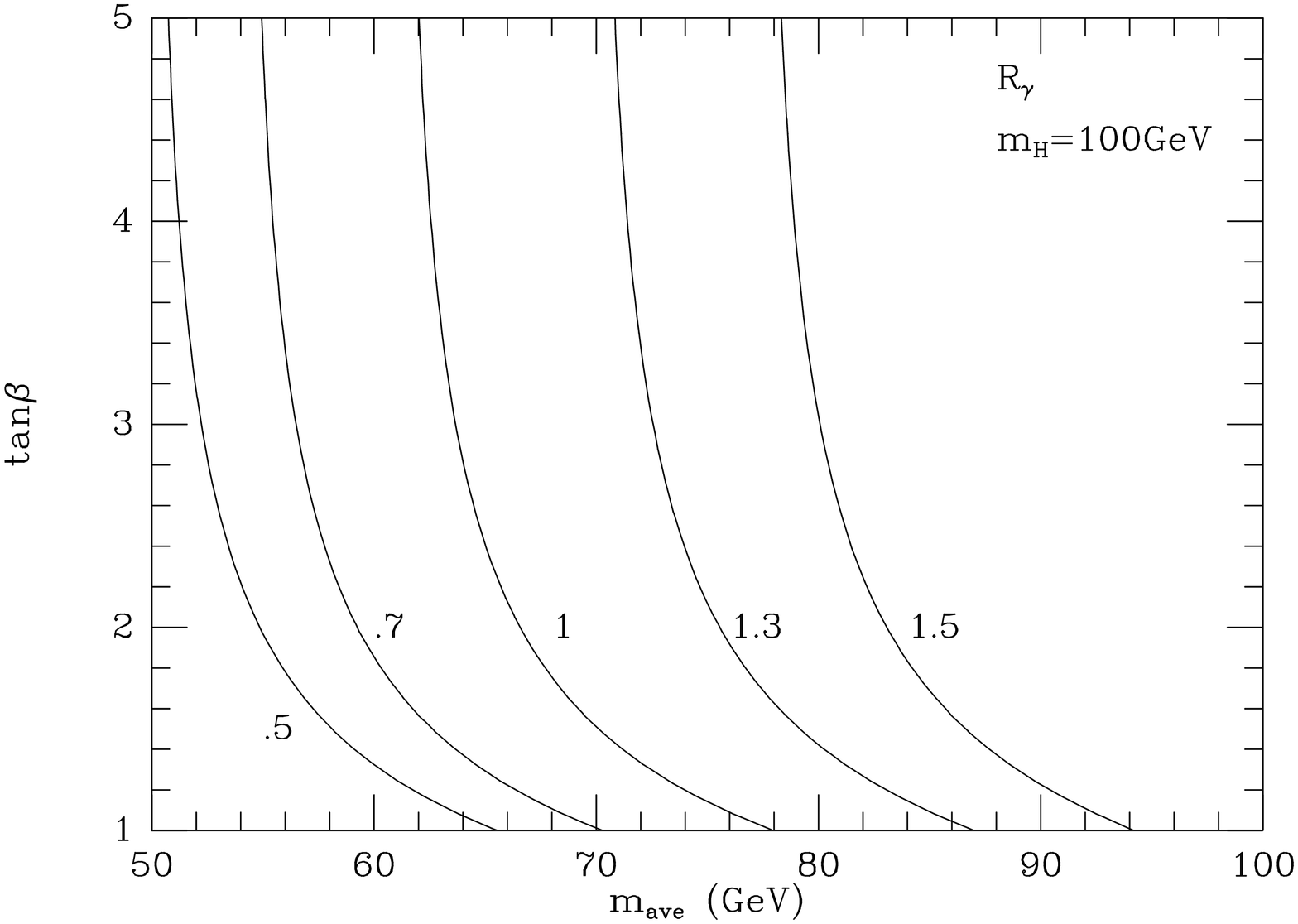,height=9cm,angle=-90}}
\vspace{0.1cm}
\centerline{
\epsfig{figure=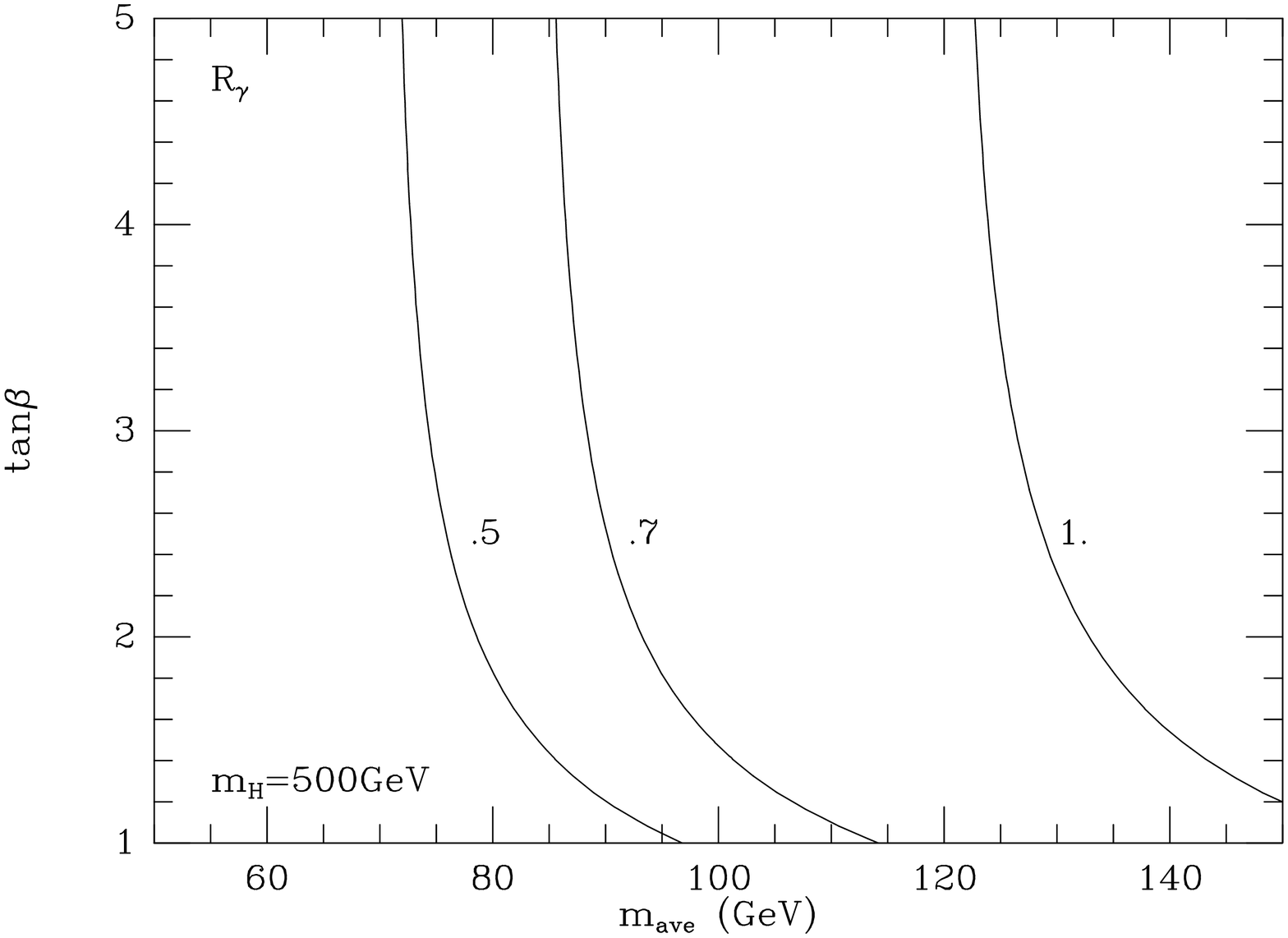,height=9cm,angle=-90}}
\vspace{0.1cm}
\caption{{\it Contours of $R_{\gamma}$ in the
$(m_{ave},\tb)$ plane, for $m_H = 100,500$ GeV.}}
\label{fig3due}
\end{figure}
\newpage
\begin{figure}[h]
\centerline{
\epsfig{figure=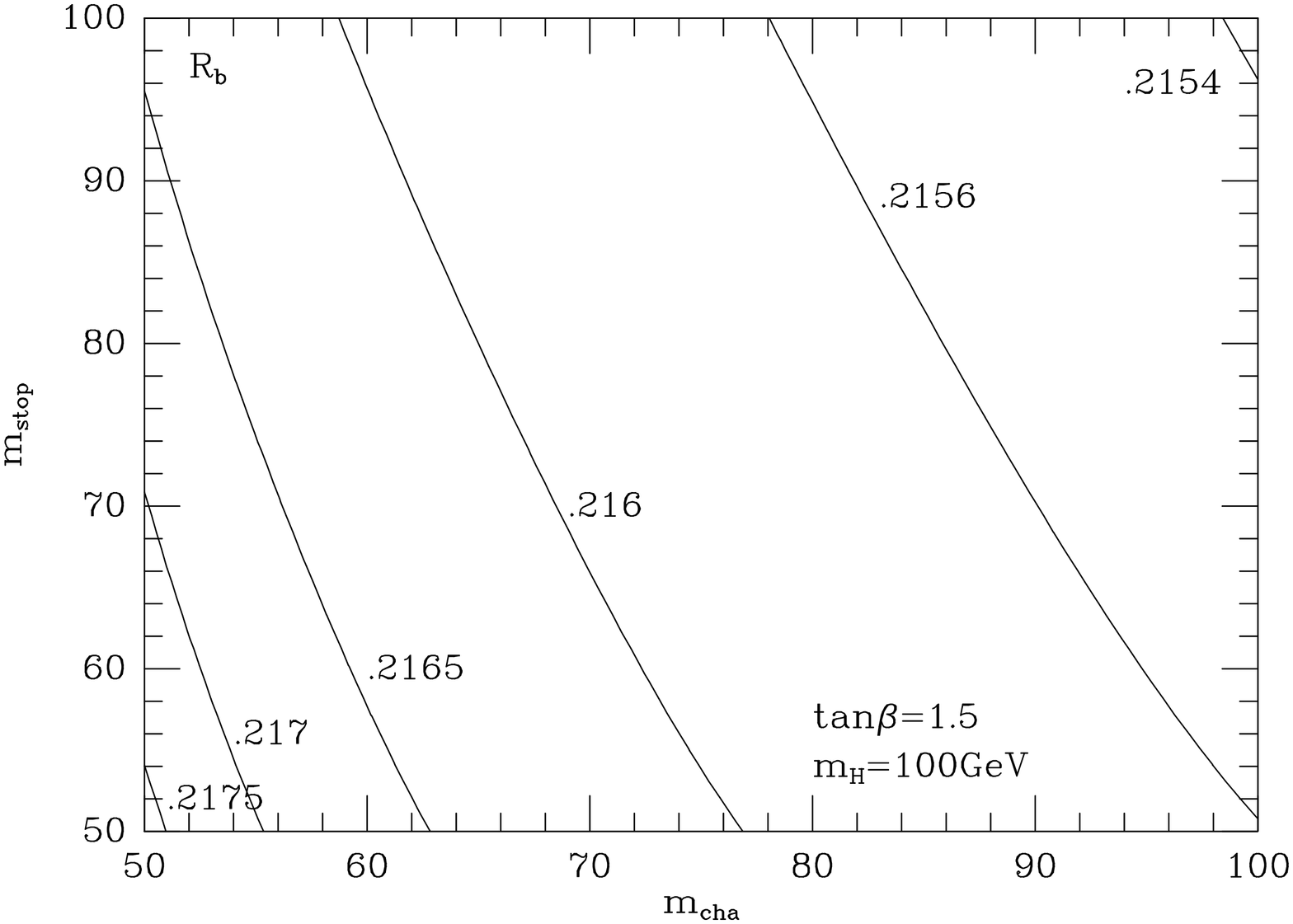,height=6cm,angle=-90}}
\vspace{-0.7cm}
\centerline{
\epsfig{figure=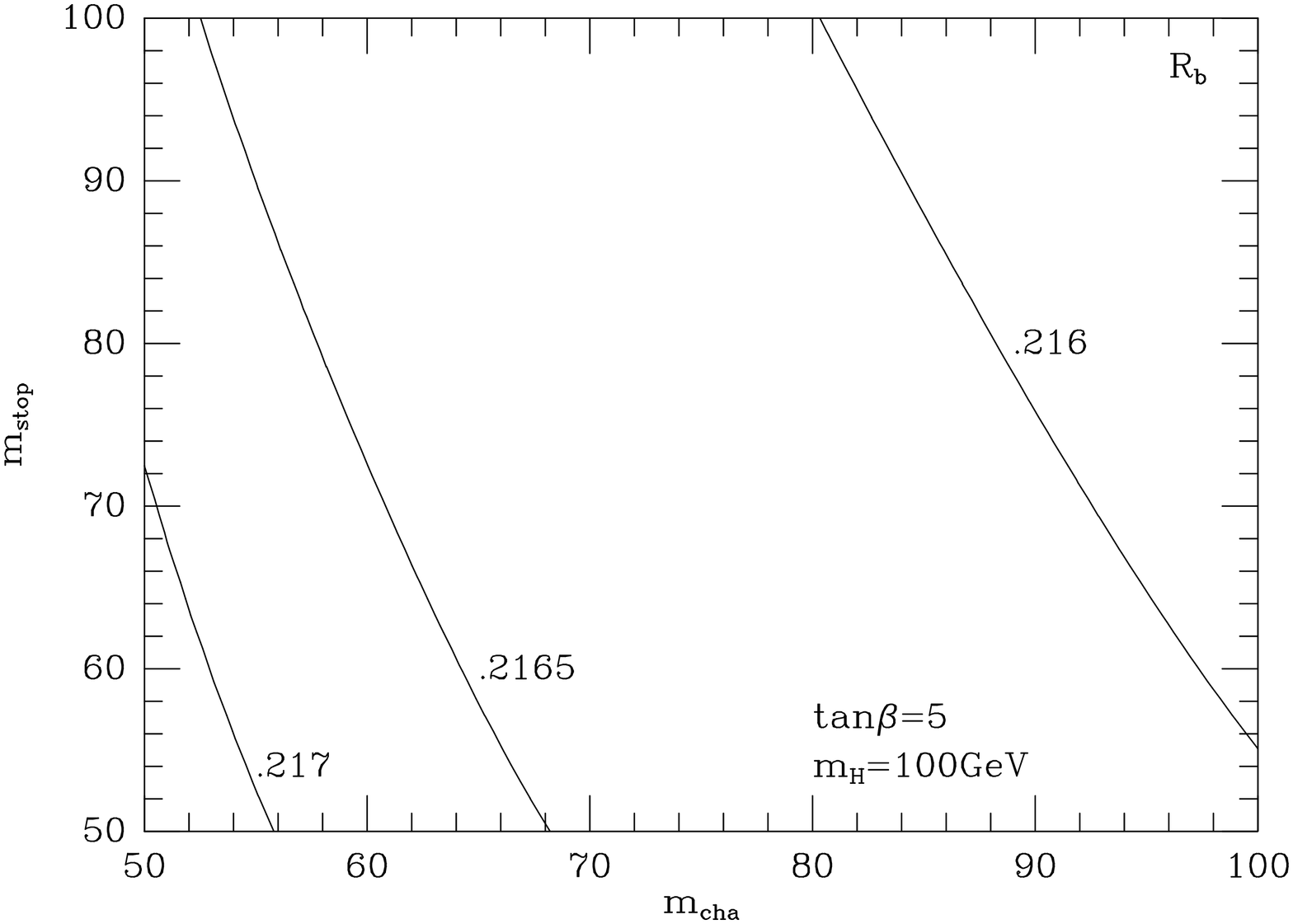,height=6cm,angle=-90}}
\vspace{-0.7cm}
\centerline{
\epsfig{figure=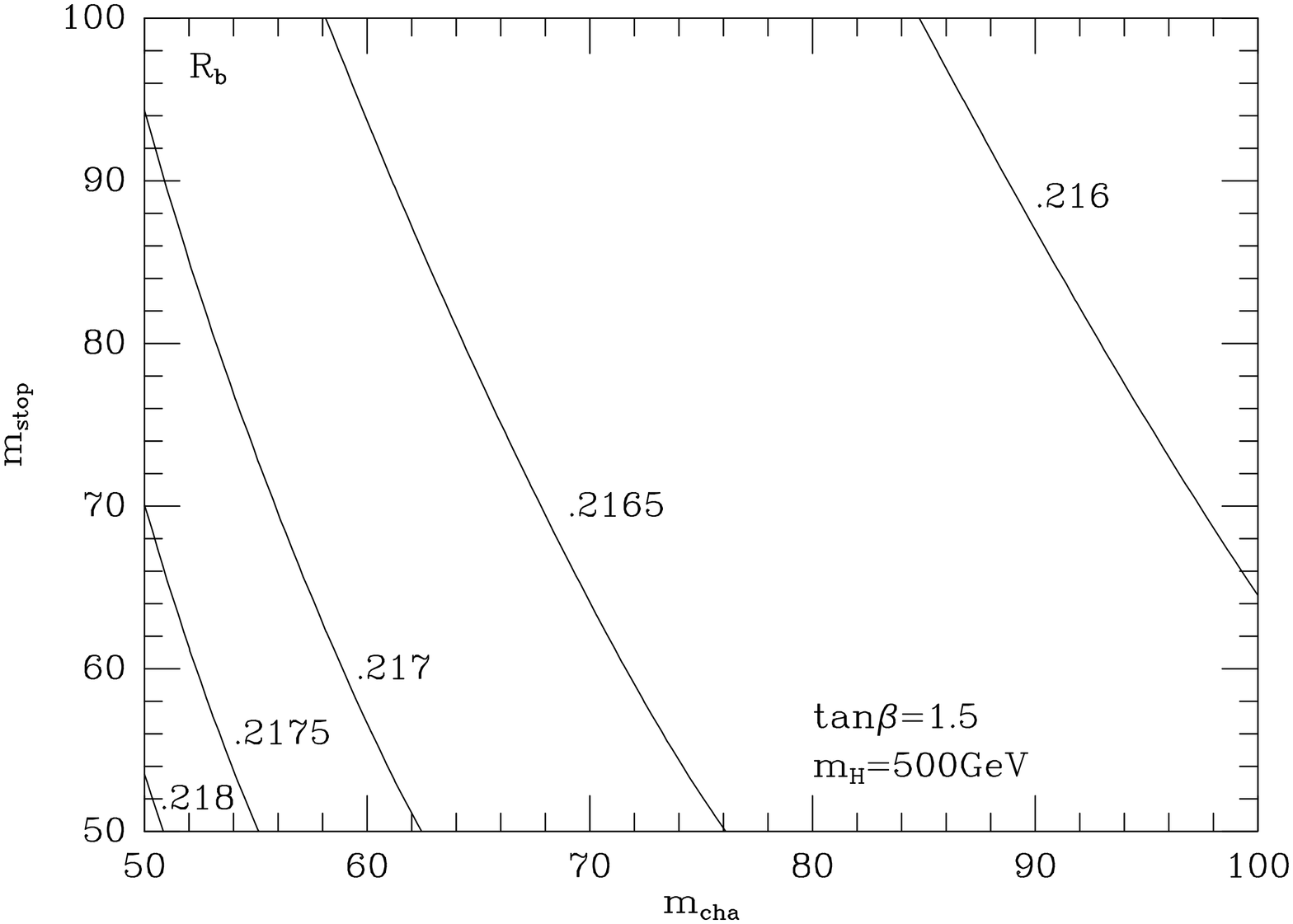,height=6cm,angle=-90}}
\vspace{-0.7cm}
\centerline{
\epsfig{figure=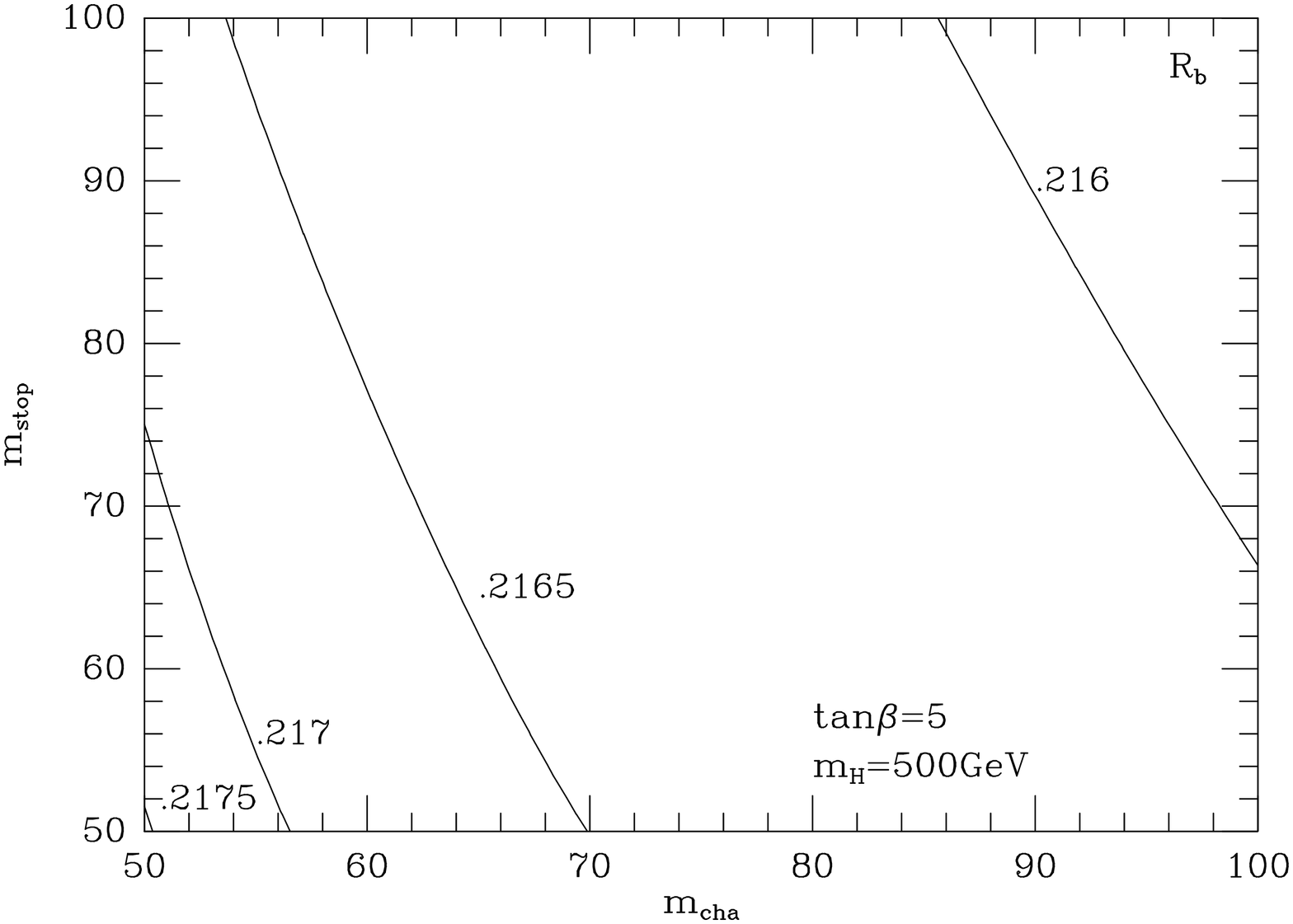,height=6cm,angle=-90}}
\vspace{-0.7cm}
\caption{\it Contours of $R_b$ in the
$(\mch,\mst)$ plane, for some representative
values of $\tb$ and $m_H$.}
\label{fig5}
\end{figure}
\newpage
\begin{figure}[h]
\centerline{
\epsfig{figure=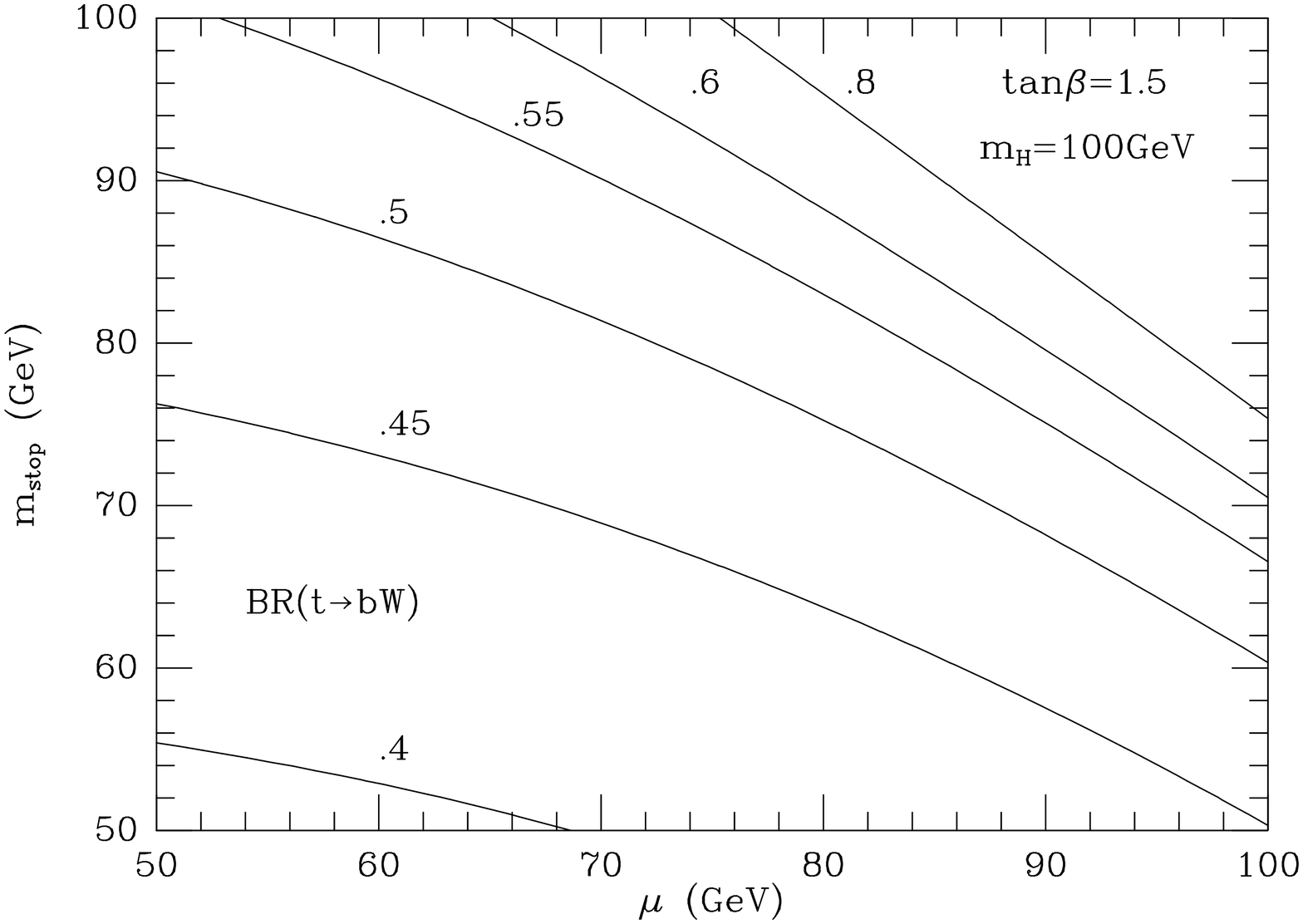,height=6cm,angle=-90}}
\vspace{-0.7cm}
\centerline{
\epsfig{figure=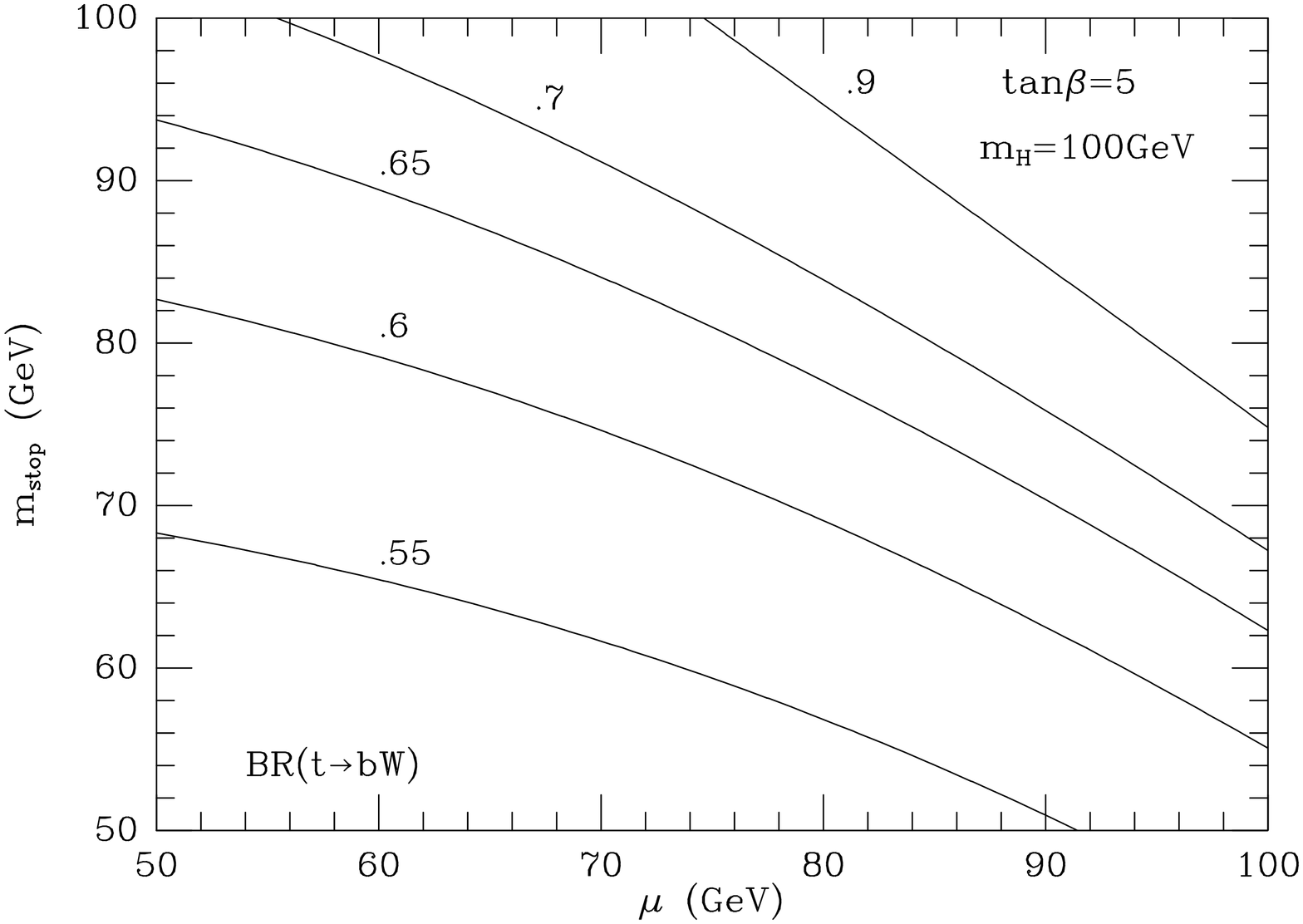,height=6cm,angle=-90}}
\vspace{-0.7cm}
\centerline{
\epsfig{figure=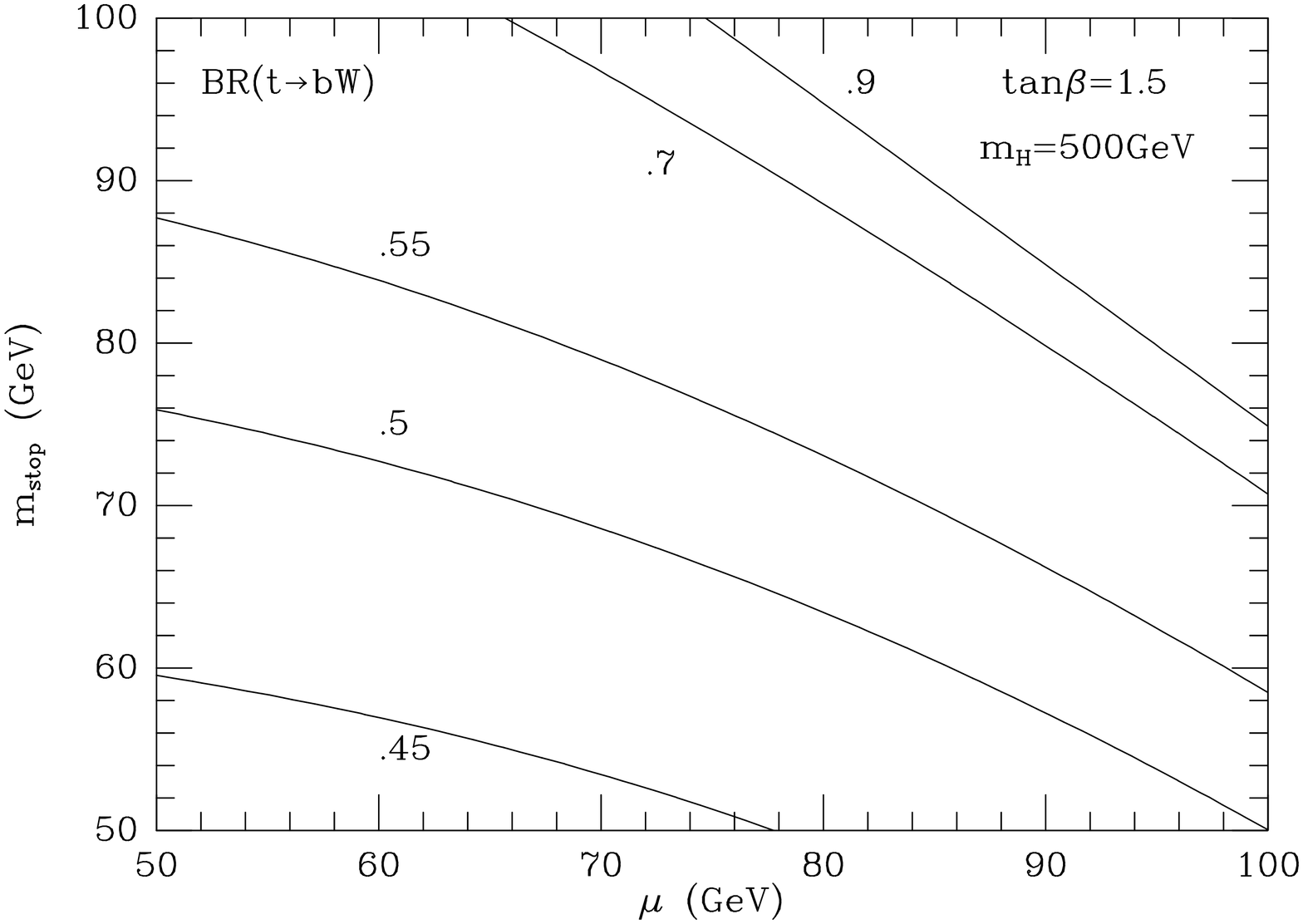,height=6cm,angle=-90}}
\vspace{-0.7cm}
\centerline{
\epsfig{figure=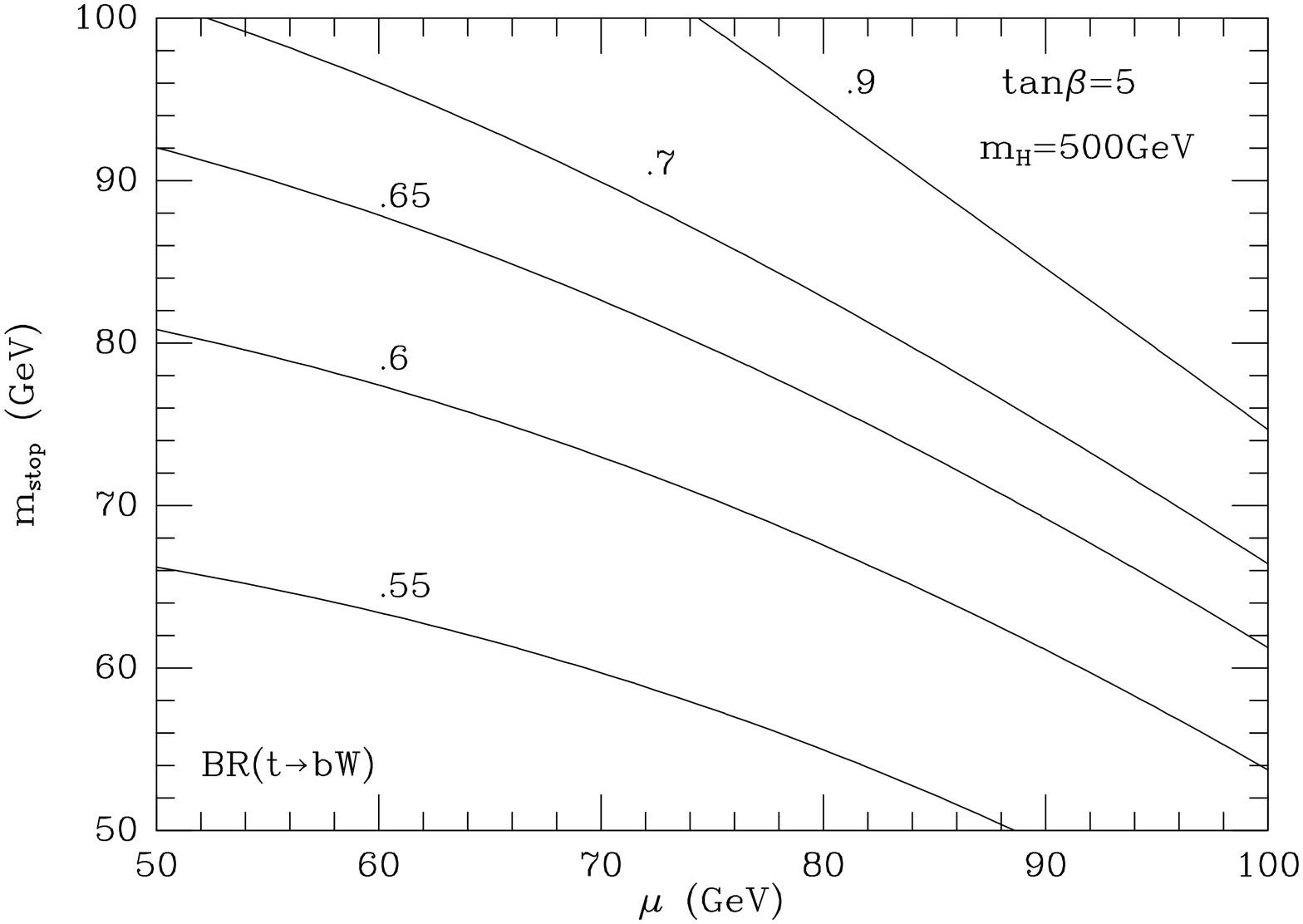,height=6cm,angle=-90}}
\vspace{-0.7cm}
\caption{\it Contours of $BR(t \rightarrow b W^+)$
in the $(\mch,\mst)$ plane, for some representative
values of $\tb$ and $m_H$.}
\label{fig4}
\end{figure}
\end{document}